\documentclass[preprint,amsmath,amssymb,aps,pra,longbibliography]{revtex4-1}

\usepackage[margin=0.75in, top=1in, bottom=0.85in]{geometry}

\usepackage[final,letterspace=-0]{microtype}
\usepackage[colorlinks=true, linkcolor=blue, urlcolor=blue, citecolor=blue, anchorcolor=blue]{hyperref}
\usepackage{mathpazo} 
\usepackage{upgreek}

\usepackage{graphicx}
\usepackage{dcolumn}
\usepackage{bm}
\usepackage{helvet}

\begin{document}

\title{Anomalous refraction of optical space-time wave packets}

\author{Basanta Bhaduri, Murat Yessenov, and Ayman F. Abouraddy}
\email{Corresponding author: raddy@creol.ucf.edu}
\affiliation{CREOL, The College of Optics \& Photonics, University of Central Florida, Orlando, FL 32816, USA}

\begin{abstract}
Refraction at the interface between two materials is fundamental to the interaction of light with photonic devices and to the propagation of light through the atmosphere at large. Underpinning the traditional rules for the refraction of an optical field is the tacit presumption of the separability of its spatial and temporal degrees-of-freedom. We show here that endowing a pulsed beam with precise spatio-temporal spectral correlations unveils remarkable refractory phenomena, such as group-velocity invariance with respect to the refractive index, group-delay cancellation, anomalous group-velocity increase in higher-index materials, and tunable group velocity by varying the angle of incidence. A law of refraction for `space-time' wave packets encompassing these effects is verified experimentally in a variety of optical materials. Space-time refraction defies our expectations derived from Fermat's principle and offers new opportunities for molding the flow of light and other wave phenomena.
\end{abstract}

\maketitle

\noindent
Snell's law, which describes the refraction of light across the interface between two media of different refractive indices, is one of the oldest principles in optics \cite{Sabra81Book}. Because of its fundamental nature, Snell's law lies at the heart of such disparate realms as the propagation of light through the atmosphere and the construction of optical instruments and devices. Refraction at an interface is essentially a spatial phenomenon involving changes in the wave momentum while conserving energy (we restrict ourselves here to non-dipsersive lossless optical media). Although Snell's law applies -- strictly speaking -- only to monochromatic plane waves, its consequences nevertheless generally extend to pulsed beams, especially for narrow spectral bandwidths in the paraxial regime in absence of dispersion. For example, the group velocity of a pulse decreases when traveling to a high-index (non-dispersive) material and the velocity of the transmitted light is independent of the angle of incidence. Such general principles provide the framework for the operation of almost all optical technologies -- from lenses and waveguides \cite{SalehBook07} to nanophotonic structures \cite{Koenderink15Science}. 

Here we show that the perennial guiding principles associated with refraction are challenged once tight spatio-temporal spectral correlations are introduced into a pulsed beam \cite{Donnelly93PRSLA,Longhi04OE,Saari04PRE,Yessenov19PRA}, whereupon unexpected phenomena are unveiled. Indeed, by associating each spatial frequency (transverse component of the wave vector) with a single wavelength \cite{Kondakci16OE,Parker16OE,Kondakci17NP}, the changes undergone by the wave momentum across an interface extend into the time domain and produce fascinating consequences that we investigate theoretically and verify experimentally. First, for any pair of materials -- regardless of their index contrast -- we find that there exists a wave packet that traverses the interface between them \textit{without changing its group velocity}, and another that retains the magnitude of its group velocity \textit{while switching sign} (the group velocity refers to the speed of the peak of the wave packet \cite{SaariPRA18}. The latter wave packet thus experiences -- surprisingly -- group-delay cancellation upon traversing equal lengths of the two materials. Second, we show that the group velocity of a wave packet can anomalously \textit{increase} when traveling from a low-index to a high-index material. Third, the group velocity of the transmitted wave packet is found to depend on the angle of incidence at the interface -- unlike the refraction of traditional wave packets. This striking effect can be exploited in synchronizing receivers at \textit{a priori unknown} locations at \textit{different} distances beyond an interface using the \textit{same} wave packet. Such unusual consequences of spatio-temporal refraction call into question our intuitions derived from Fermat's principle, which undeniably governs each underlying monochromatic plane wave but does not extend to the wave packet as a whole once endowed with tight spatio-temporal spectral correlations. These predictions are verified through interferometric group-delay measurements in a variety of optical materials.

The spectral loci of these `space-time' (ST) wave packets on the surface of the light-cone are confined to reduced-dimensionality trajectories with respect to traditional pulsed beams \cite{Donnelly93PRSLA,Yessenov19PRA}. The reduced dimensionality of the spectral representation is a consequence of associating each spatial frequency with a single wavelength, in contradistinction to traditional wave packets in which the spatial and temporal spectra are separable, such that each spatial frequency is associated with a finite bandwidth \cite{Kondakci17NP}. When the spectral trajectory lies at the intersection of the light-cone with a tilted spectral plane \cite{Donnelly93PRSLA,Kondakci17NP}, the ST wave packet is transported rigidly \cite{Besieris89JMP,Saari97PRL,Salo01JOA,Turunen10PO,FigueroaBook14,Porras17OL,Wong17ACSP2,Efremidis17OL} at a group velocity dictated solely by the spectral tilt angle of this plane \cite{Kondakci19NC} independently of the refractive index \cite{Bhaduri19Optica}. The spectral tilt angle $\theta$ of a ST wave packet is an \textit{internal} degree of freedom that characterizes the global properties of the field independently of its \textit{extrinsic} degrees of freedom (such as central wavelength, bandwidth, beam size and profile, or direction of propagation). By identifying a quantity characteristic of the global properties of the ST wave packet that is invariant after traversing a planar interface, we formulate an expression for the change in the spectral tilt angle and hence the group velocity upon refraction. Whereas Snell's law governs an external degree of freedom (the propagation angle), the expression we derive governs an internal degree of freedom (the spectral tilt angle), and thus represents a new law of refraction unique to ST wave packets.

\begin{figure*}[t!]
\centering
\includegraphics[width=17.6cm]{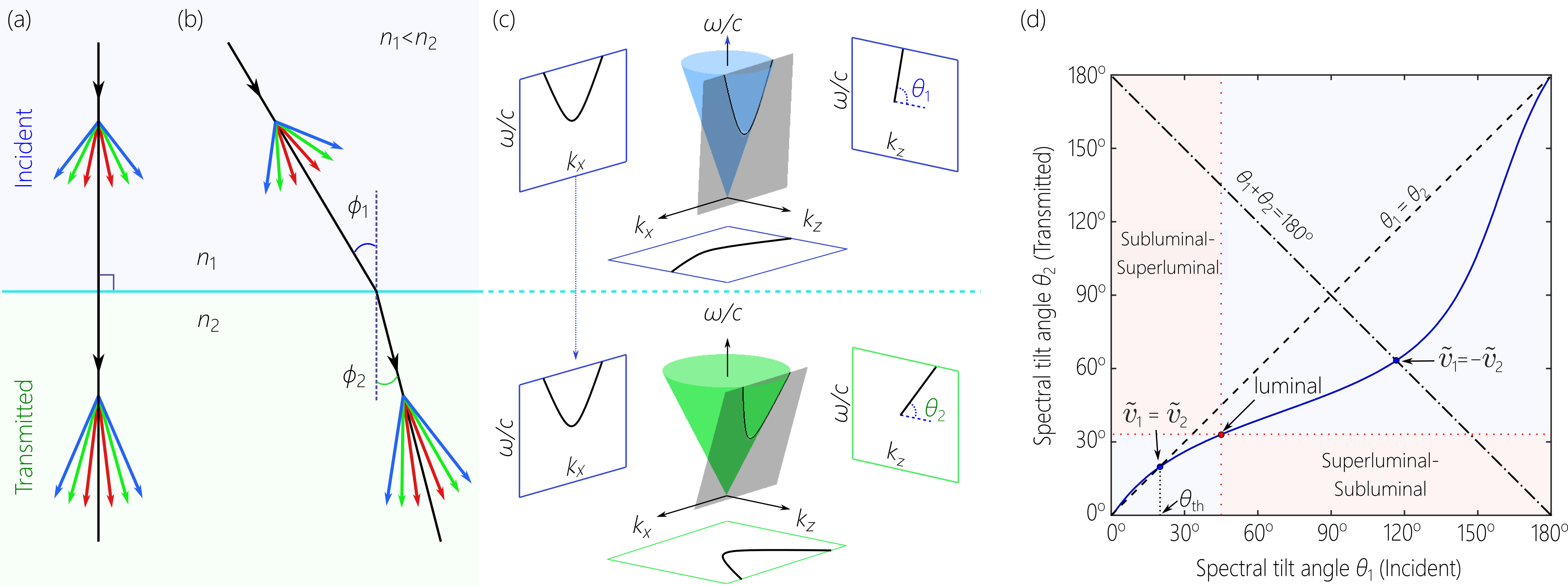}
\caption{Dynamical refraction of ST wave packets. (a) An ST wave packet is incident normally and (b) obliquely at the interface between two semi-infinite optical materials. (c) The spatio-temporal spectrum of the ST wave packet in the first material lies along the intersection of the light-cone (apex angle $\tan^{-1}{n_{1}}$) with a spectral hyperplane having a spectral tilt angle $\theta_{1}$. For normal incidence, the projection onto the $(k_{x},\tfrac{\omega}{c})$-plane is invariant, enforcing the spectral tilt angle in the second material (light-cone apex angle $\tan^{-1}{n_{2}}$) to take on a new value $\theta_{2}$. (d) The relationship between $\theta_{1}$ and $\theta_{2}$ based on Eq.~\ref{Eq:SnellsLawBaseband}. The overall features of the curve are generic, but for concreteness we used $n_{1}\!=\!1$ and $n_{2}\!=\!1.5$.}
\label{Fig:Concept}
\end{figure*}

We start by examining the refraction of a ST wave packet at normal incidence on a planar interface between two semi-infinite, non-dispersive, isotropic, homogeneous materials of refractive indices $n_{1}$ and $n_{2}$ (Fig.~\ref{Fig:Concept}a). In a material of refractive index $n$, the optical field can be expanded into monochromatic plane waves $e^{i(k_{x}x+k_{z}z-\omega t)}$, each represented by a point on the surface of the light-cone $k_{x}^{2}+k_{z}^{2}\!=\!(n\tfrac{\omega}{c})^{2}$. Here $k_{x}$ and $k_{z}$ are the transverse and axial components of the wave vector along $x$ and $z$, respectively, $\omega$ is the temporal frequency, and for simplicity we hold the field uniform along $y$. The spatio-temporal spectral representation of a typical pulsed beam occupies a two-dimensional domain on the light-cone surface \cite{Kondakci17NP}. We consider here propagation-invariant ST wave packets whose representations lie along one-dimensional curved trajectories (conic sections) \cite{Faccio07OE} at the intersection of the light-cone with tilted spectral planes $\tfrac{\omega}{c}\!=\!k_{\mathrm{o}}+(k_{z}-nk_{\mathrm{o}})\tan{\theta}$, where $k_{\mathrm{o}}$ is a fixed wave number and $\theta$ is the spectral tilt angle with respect to the $k_{z}$-axis (Fig.~\ref{Fig:Concept}c) \cite{Donnelly93PRSLA,Kondakci17NP}. This internal degree of freedom $\theta$ solely dictates the group velocity $\widetilde{v}\!=\!c\tan{\theta}\!=\!c/\widetilde{n}$, where $\widetilde{n}\!=\!\cot{\theta}$ is the group index. The \textit{sub}luminal regime corresponds to $\widetilde{v}\!<\!c/n$ ($\widetilde{n}\!>\!n$), and the \textit{super}luminal to $\widetilde{v}\!>\!c/n$ ($\widetilde{n}\!<\!n$). Such wave packets offer uncommon flexibility for tuning $\widetilde{v}$ in free space \cite{Kondakci19NC} and non-dispersive materials \cite{Bhaduri19Optica}; see Supplementary. 

\begin{figure}[t!]
\centering
\includegraphics[width=8.6cm]{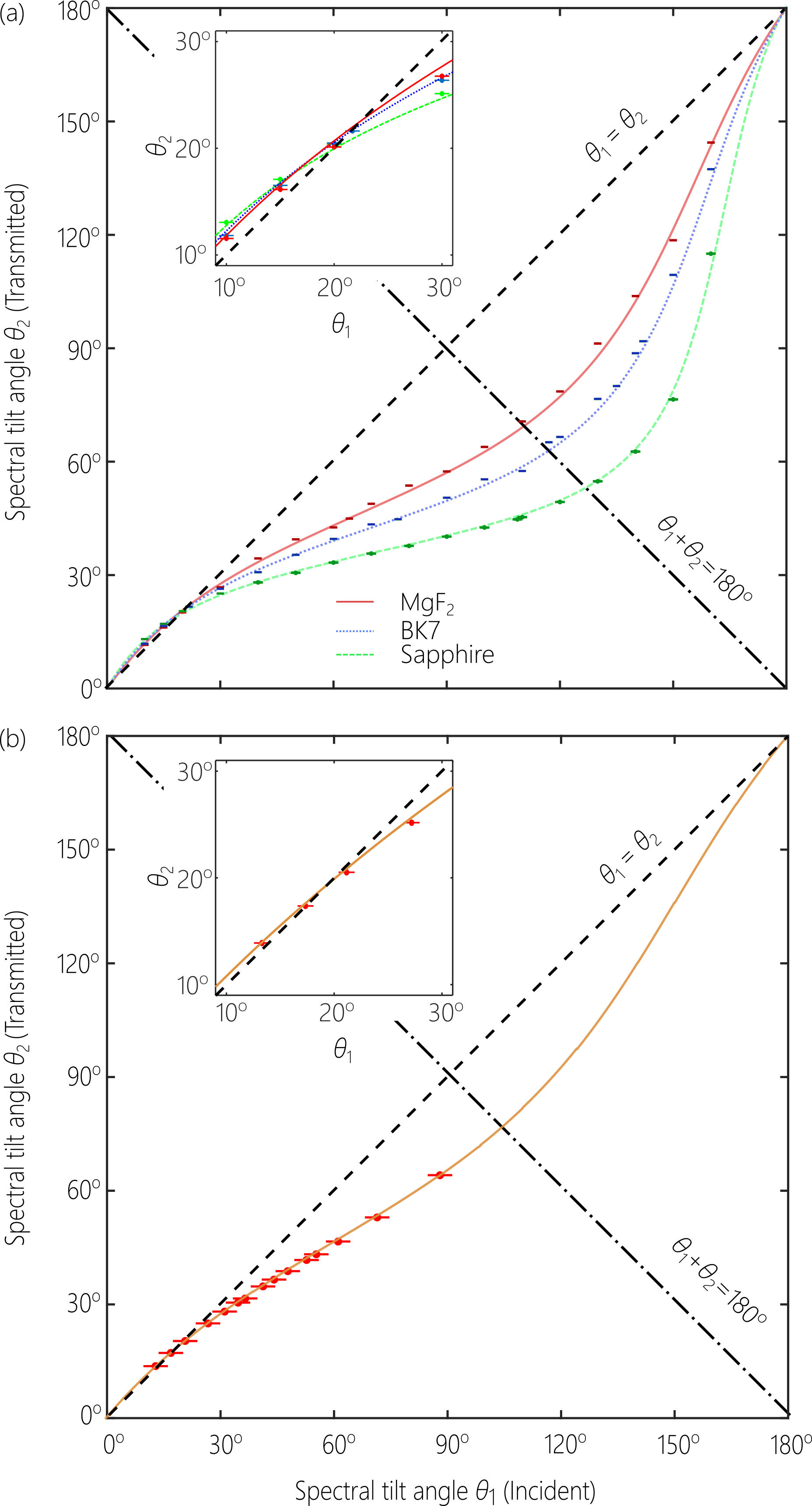}
\caption{(a) Experimental verification of the law of refraction of ST wave packets at normal incidence from free space onto MgF$_2$, BK7 glass, and sapphire; and (b) from BK7 to sapphire. All materials are in the form of 5-mm-thick windows. The points are data and the curves correspond to Eq.~\ref{Eq:SnellsLawBaseband}. The insets highlight the anomalous refraction regime.}
\label{Fig:Snell}
\end{figure}

The light-cone angle changes with $n$, so that the transition from one medium to another leads to a diffeomorphism of the ST wave-packet representation constrained by the invariance of $\omega$ (conservation of energy) and $k_{x}$ (conservation of transverse momentum due to shift-invariance along $x$) across a planar interface at normal incidence; see Fig.~\ref{Fig:Concept}c. Approximating the conic section representing the spatio-temporal spectral trajectory of the wave packet on the light-cone by a parabola at small bandwidths (with respect to the central frequency; see Supplementary) \cite{Kondakci17NP,Kondakci19NC,Bhaduri19Optica}, we identify the quantity $n(n-\widetilde{n})$ that is proportional to the curvature of the spectral representation as an \textit{invariant} at normal incidence, thus leading to the following law of refraction for ST wave packets:
\begin{equation}\label{Eq:SnellsLawBaseband}
n_{1}(n_{1}-\widetilde{n}_{1})=n_{2}(n_{2}-\widetilde{n}_{2}),
\end{equation}
where $n_{1}$ and $n_{2}$ are the refractive indices of the two materials, and $\widetilde{n}_{1}$ and $\widetilde{n}_{2}$ are the group indices of the incident and transmitted fields, respectively. We plot in Fig.~\ref{Fig:Concept}d the formula in Eq.~\ref{Eq:SnellsLawBaseband} in terms of the spectral tilt angles $\theta_{1}$ and $\theta_{2}$ when $n_{1}\!<\!n_{2}$. Equivalently, this transformation can be plotted between the group indices or the group velocities of the incident and transmitted wave packets (Supplementary). We verify the law of refraction in Eq.~\ref{Eq:SnellsLawBaseband} utilizing ST wave packets of width $\sim$9~ps at a wavelength of $\sim$800~nm. The narrow bandwidth associated with these wave packets helps avoid any effects of dispersion. Relying on interferometric measurements, we determine the group delay incurred by the ST wave packet (whose free-space group velocity can be tuned continuously \cite{Kondakci19NC}) in comparison to the group delay of a generic laser pulse when both traverse a layer or bilayer of optical materials, from which we can extract the wave packet spectral tilt angle in each material (Supplementary) \cite{Kondakci19NC,Bhaduri19Optica}. We trace out in Fig.~\ref{Fig:Snell}a the law of refraction at normal incidence from free space ($n_{1}\!=\!1$) onto MgF$_2$ ($n_{2}\!\approx\!1.38$), BK7 glass ($n_{2}\!\approx\!1.51$), and sapphire ($n_{2}\!\approx\!1.76$), in addition to the interface between BK7 and sapphire (Fig.~\ref{Fig:Snell}b). This law is independent of the external degrees of freedom of the field and applies regardless of the details of the transverse beam profile or temporal pulse linewidth (Supplementary). Fresnel reflection at the surface may alter the spatio-temporal spectral amplitudes, thereby potentially changing the profile of the transmitted wave packet, but does not affect the change in group velocity as predicted by Eq.~\ref{Eq:SnellsLawBaseband}.

Despite its simplicity, the formula in Eq.~\ref{Eq:SnellsLawBaseband} has far-reaching consequences. An immediate result is that the subluminal-to-superluminal barrier cannot be crossed by traversing an interface: a subluminal ST wave packet $\widetilde{n}_{1}\!>\!n_{1}$ (superluminal $\widetilde{n}_{1}\!<\!n_{1}$) in the first material remains subluminal $\widetilde{n}_{2}\!>\!n_{2}$ (superluminal $\widetilde{n}_{2}\!<\!n_{2}$) in the second. We pose the following question: can the group index of a ST wave packet remain invariant ($\widetilde{n}_{1}\!=\!\widetilde{n}_{2}$) upon traversing the interface? Equation~\ref{Eq:SnellsLawBaseband} indicates that this can indeed occur in the subluminal regime at a threshold group index $\widetilde{n}_{\mathrm{th}}\!=\!n_{1}+n_{2}$, whereupon $\widetilde{n}_{1}\!=\!\widetilde{n}_{2}$ and $\widetilde{v}_{1}\!=\!\widetilde{v}_{2}$.
This threshold separates `normal' and `anomalous' refraction regimes. In the normal-refraction regime $\widetilde{n}_{1}\!<\!\widetilde{n}_{\mathrm{th}}$, the group velocity of the transmitted wave packet drops $\widetilde{v}_{2}\!<\!\widetilde{v}_{1}$ as usual when $n_{1}\!<\!n_{2}$. In contrast, in the anomalous-refraction regime $\widetilde{n}_{1}\!>\!\widetilde{n}_{\mathrm{th}}$, the group velocity counter-intuitively increases $\widetilde{v}_{2}\!>\widetilde{v}_{1}$ despite the higher refractive index. Previous theoretical studies examined the refraction of focus-wave modes \cite{Hillion93Optik,Donnelly97IEEE} and X-waves \cite{Attiya01PER,Salem12JOSAA} whose velocities are restricted to superluminal values $\widetilde{v}\!>\!c$ \cite{Saari97PRL}, and thus do not display the effects uncovered here that occur necessarily in the subluminal regime.

\begin{figure}[t!]
\centering
\includegraphics[width=8.6cm]{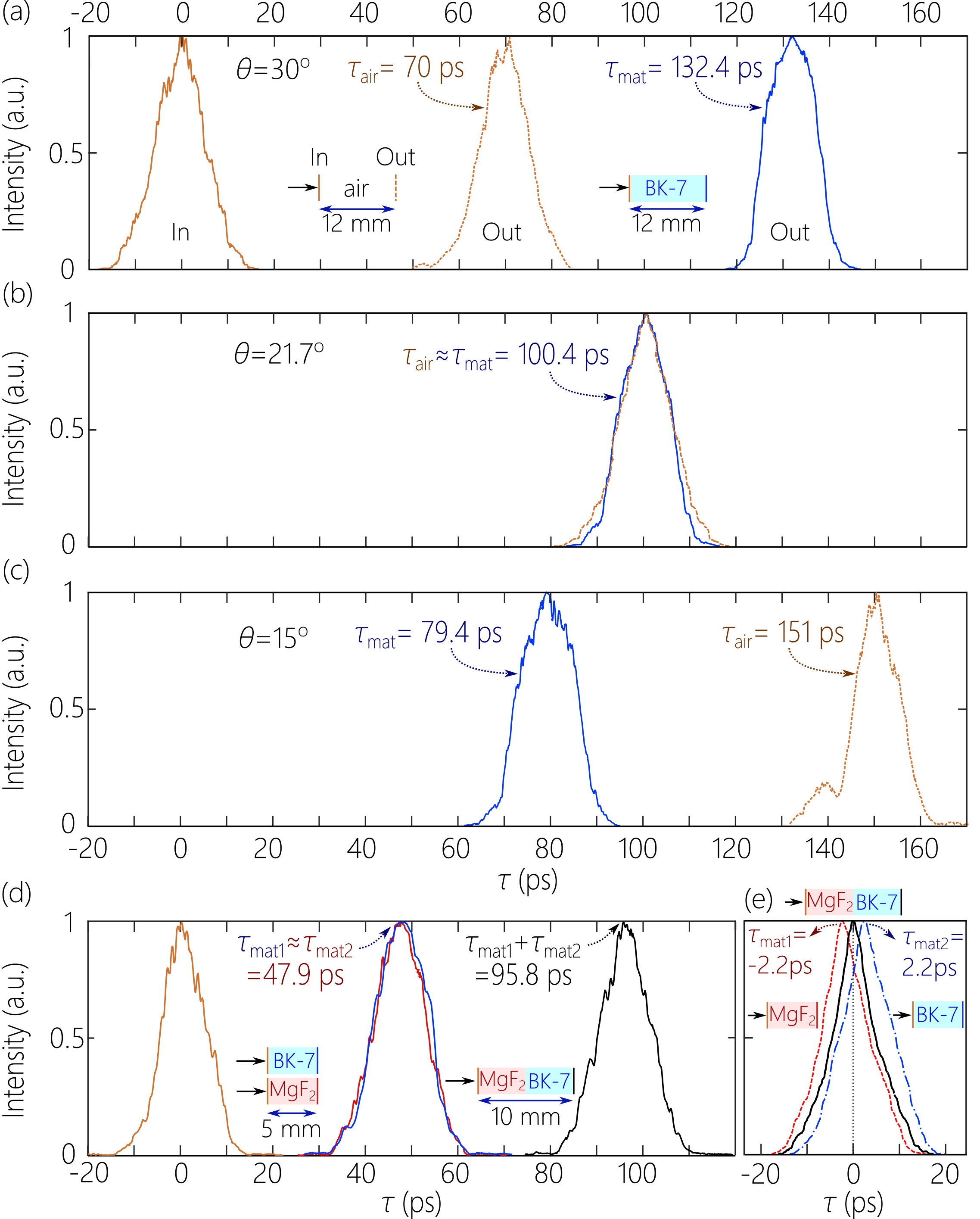}
\caption{Confirmation of the predictions of the law of refraction for ST wave packets at normal incidence (Eq.~\ref{Eq:SnellsLawBaseband}). (a-c) Temporal envelope of a ST wave packet at the center of the spatial profile after traversing $L\!=\!12$~mm in air (incurring a delay $\tau_{\mathrm{air}}$; dotted brown curve) and in BK7 (incurring a delay $\tau_{\mathrm{mat}}$; solid blue curve) while changing the spectral tilt angle $\theta$ of the incident wave packet. (a) At $\theta_{1}\!=\!30^{\circ}$ in the normal-refraction regime, $\tau_{\mathrm{mat}}\!>\!\tau_{\mathrm{air}}$; the wave packet is slower in BK7 than in air. The solid brown curve on the left is the incident ST wave packet. (b) At the threshold $\theta_{1}\!=\!21.7^{\circ}$, $\tau_{\mathrm{mat}}\!=\!\tau_{\mathrm{air}}$; the wave packet travels in air and in BK7 at the same velocity. (c) At $\theta_{1}\!=\!15^{\circ}$ in the anomalous-refraction regime $\tau_{\mathrm{mat}}\!<\!\tau_{\mathrm{air}}$; the wave packet travels in BK7 faster than in air. (d) Refraction at the threshold $\widetilde{n}_{\mathrm{th}}\!=\!2.89$ ($\theta_{\mathrm{th}}\!\approx\!19^{\circ}$) for MgF$_2$ and BK7 ($L\!=\!5$~mm for each material). The group delays $\tau_{\mathrm{mat}1}$ and $\tau_{\mathrm{mat}2}$ in the two materials are equal, and the group delay in a bilayer is double that of a single layer. (e) Group-delay cancellation in a bilayer of equal lengths ($L\!=\!5$~mm each) of MgF$_2$ and BK7. The group delay $\tau_{\mathrm{mat}1}$ in MgF$_2$ is positive whereas the group delay $\tau_{\mathrm{mat}2}$ in BK7 is negative, with $\tau_{\mathrm{mat}1}\!=\!-\tau_{\mathrm{mat}2}\!\approx\!2.2$~ps so that the total delay in the bilayer is $\tau_{\mathrm{mat}1}+\tau_{\mathrm{mat}2}\!=\!0$. This condition corresponds to a free-space spectral-tilt-angle of $\theta\!=\!137.1^{\circ}$ (Supplementary).}
\label{Fig:Measurements}
\end{figure}

We verify normal and anomalous refraction at the interface between free space and BK7 where $\widetilde{n}_{\mathrm{th}}\!=\!2.51$ ($\theta_{\mathrm{th}}\!=\!21.7^{\circ}$). In Fig.~\ref{Fig:Measurements}a-c we plot the temporal envelope of a ST wave packet after traversing $L\!=\!12$~mm of air (where it accrues a group delay $\tau_{\mathrm{air}}$) and of BK7 (group delay $\tau_{\mathrm{mat}}$). At $\theta_{1}\!=\!30^{\circ}\!>\!\theta_{\mathrm{th}}$ ($\widetilde{n}_{1}\!=\!1.73\!<\!\widetilde{n}_{\mathrm{th}}$) in the normal refraction regime we have $\tau_{\mathrm{mat}}\!>\!\tau_{\mathrm{air}}$ as usual (Fig.~\ref{Fig:Measurements}a); the group velocity is lower in the higher-index BK7 with respect to air. Reducing $\theta_{1}$ to $\theta_{\mathrm{th}}$ results in $\tau_{\mathrm{mat}}\!=\!\tau_{\mathrm{air}}$, indicating that $\widetilde{v}_{1}\!=\!\widetilde{v}_{2}$ at the threshold (Fig.~\ref{Fig:Measurements}b); the wave-packet group velocity is the same in air and in BK7. By further reduction to $\theta_{1}\!=\!15^{\circ}\!<\!\theta_{\mathrm{th}}$ ($\widetilde{n}_{1}\!=\!3.73\!>\!\widetilde{n}_{\mathrm{th}}$) in the anomalous refraction regime, we have $\tau_{\mathrm{mat}}\!<\!\tau_{\mathrm{air}}$ (Fig.~\ref{Fig:Measurements}c), indicating that $\widetilde{v}_{1}\!<\!\widetilde{v}_{2}$; anomalously, the wave packet has a higher group velocity in BK7 than in air. Furthermore, we confirm in Fig.~\ref{Fig:Measurements}d the threshold condition at the interface between MgF$_2$ and BK7 when $\theta_{\mathrm{th}}\!\approx\!19^{\circ}$ and $\widetilde{n}_{\mathrm{th}}\!=\!2.89$ in both materials (corresponding to $\theta\!\approx\!18^{\circ}$ in free space). The group delay is equal in $L\!=\!5$~mm of either material, and is doubled in a bilayer of them. 

These predictions are all the more counter-intuitive from the standpoint of the spectral representation of the field on the light-cone (Supplementary). Because the light-cone angle increases with $n$, the surface of the light-cone inflates in a medium with higher $n$ (we assume here that $n_{2}\!>\!n_{1}$). The surprising nature of anomalous refraction is best grasped by examining the spectral projection onto the $(k_{z},\tfrac{\omega}{c})$-plane. Conservation of energy and transverse momentum dictate that the widths of the spectral projections along the $\tfrac{\omega}{c}$ and $k_{x}$ axes are fixed; which we denote $\tfrac{\Delta\omega}{c}$ and $\Delta k_{x}$, referring to the temporal and spatial bandwidths, respectively. Traditionally, the light-cone inflation with $n$ together with the invariance of the temporal bandwidth $\tfrac{\Delta\omega}{c}$ lead to an increase in the projection along the $k_{z}$-axis ($\Delta k_{z}$) and therefore a reduction in the slope of the spectral projection onto the $(k_{z},\tfrac{\omega}{c})$-plane, $\widetilde{n}\!=\!\tfrac{\Delta k_{z}}{\Delta\omega/c}$; hence the familiar reduction in $\widetilde{v}$ in higher-index non-dispersive media. At first glance, it seems that spatio-temporal spectral structuring cannot circumvent this constraint. However, the reduced-dimensionality of the spectral representation of ST wave packets reveals a geometric effect that is concealed when considering traditional pulses. Indeed, the invariant temporal and spatial bandwidths that are tightly correlated combine to \textit{shrink} the projection $\Delta k_{z}$ along the $k_{z}$-axis with increasing $n$. It can be shown that the slope of the spectral projection onto the $(k_{z},\tfrac{\omega}{c})$-plane after combining both effects is
\begin{equation}\label{eq:interplay}
\widetilde{n}=\frac{\Delta k_{z}}{\Delta\omega/c}\approx\,\, n+\frac{1}{2n}\left(\frac{\Delta k_{x}}{\Delta\omega/c}\right)^{2};
\end{equation}
where the ratio of the spatial and temporal bandwidths in the second term is an invariant. Therefore, the group velocity of the transmitted wave packet is determined by the interplay between two opposing trends upon changing $n$: an increase in $\Delta k_{z}$ due to light-cone inflation that reduces $\widetilde{v}_{2}$ (first term in Eq.~\ref{eq:interplay}); and an opposing shrinkage in $\Delta k_{z}$ due to the invariance of the correlated bandwidths $\tfrac{\Delta\omega}{c}$ and $\Delta k_{x}$ that increases $\widetilde{v}_{2}$ (second term in Eq.~\ref{eq:interplay}, which is negligible for tradition wave packets). It can be readily shown that these two opposing effects balance each other out exactly for incident ST wave packets having a group velocity corresponding to $\widetilde{n}_{\mathrm{th}}\!=\!n_{1}+n_{2}$, in which case the group velocity remains invariant $\widetilde{v}_{1}\!=\!\widetilde{v}_{2}\!=\!c/\widetilde{n}_{\mathrm{th}}$ after traversing the interface regardless of the index contrast. In the normal-refraction regime ($\theta_{1}\!>\!\theta_{\mathrm{th}}$ or $\widetilde{n}_{1}\!<\!\widetilde{n}_{\mathrm{th}}$), the light-cone inflation dominates so that $\widetilde{v}_{1}\!>\!\widetilde{v}_{2}$; whereas in the anomalous-refraction regime ($\theta_{1}\!<\!\theta_{\mathrm{th}}$ or $\widetilde{n}_{1}\!>\!\widetilde{n}_{2}$), the constraint-induced shrinkage rate along the $k_{z}$-axis exceeds the inflation rate so that $\widetilde{v}_{1}\!<\!\widetilde{v}_{2}$.

In the superluminal regime $\widetilde{n}_{1}\!<\!n_{1}$, the group velocity always decreases when going from low to high index as with traditional pulses. However, a striking scenario occurs at the unique intersection of the curve in Fig.~\ref{Fig:Concept}d with the anti-diagonal $\theta_{1}+\theta_{2}\!=\!180^{\circ}$, whereupon $\widetilde{n}_{1}\!=\!n_{1}-n_{2}$ and $\widetilde{n}_{2}\!=\!n_{2}-n_{1}\!=\!-\widetilde{n}_{1}$; that is, the magnitude of the group velocity is constant while its sign flips $\widetilde{v}_{2}\!=\!-\widetilde{v}_{1}$, leading to \textit{cancellation} of the group delay accrued upon traversing equal lengths of these two materials. We confirm this predicted group-delay cancellation after traversing a bilayer of MgF$_2$ and BK7 ($L\!=\!5$~mm for each). We plot in Fig.~\ref{Fig:Measurements}e the ST wave packet after traversing each layer separately and then traversing the bilayer confirming that zero group delay is accrued upon traversing the pair. Our experiments have made use of generic widely used optical materials, but the results extend to all materials in absence of chromatic dispersion. 

\begin{figure}[t!]
\centering
\includegraphics[width=8.6cm]{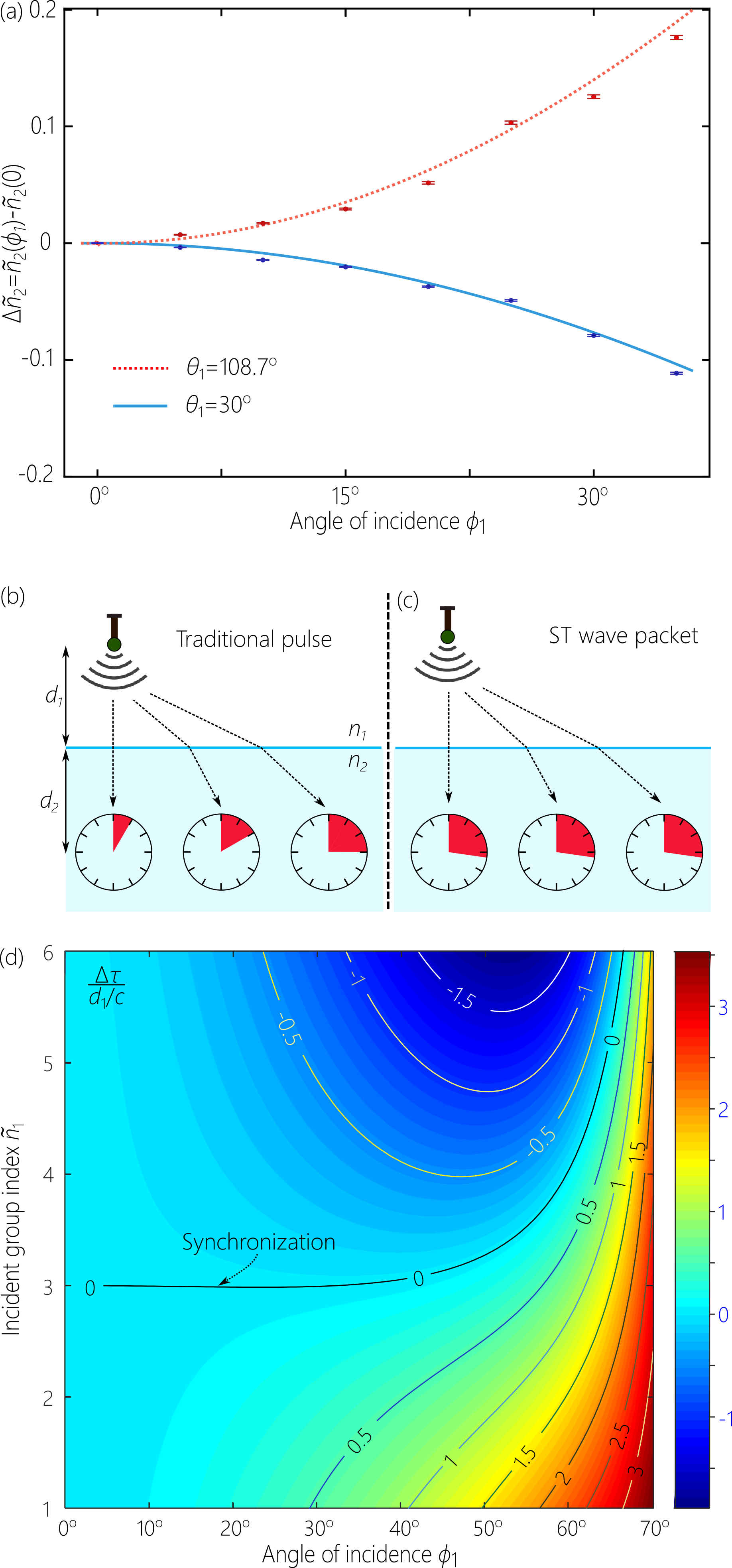}
\caption{(a) Change in the group index of the transmitted ST wave packet $\Delta\widetilde{n}_{2}(\phi_{1})\!=\!\widetilde{n}_{2}(\phi_{1})-\widetilde{n}_{2}(0)$ with incidence angle $\phi_{1}$ for subluminal ($\theta_{1}\!=\!30^{\circ}$) and superluminal ($\theta_{1}\!=\!108.7^{\circ}$) wave packets. Incidence is from air onto sapphire. Points are data and the curves are theoretical predictions based on Eq.~\ref{Eq:ObliqueSnellsLawBaseband}. (b-c) Schematic of the configuration for synchronizing remote stations utilizing (b) a traditional pulse and (c) a ST wave packet. The source is located at a distance $d_{1}$ above the interface in a medium of refractive index $n_{1}$ and the receivers are all at a depth $d_{2}$ below it in a medium of index $n_{2}$. (d) Plot of $\Delta\tau\!=\!\Delta\tau_{1}+\Delta\tau_{2}$ (normalized with respect to $d_{1}/c$) with $\phi_{1}$ and the group index $\widetilde{n}_{1}$ of a wave packet incident from air onto sapphire ($d_{2}/d_{1}\!=\!5$). Synchronization $\Delta\tau\!\approx\!0$ occurs in the angle-of-incidence range $-30^{\circ}\!<\!\phi_{1}\!<\!30^{\circ}$.}
\label{Fig:ObliqueIncidence}
\end{figure}

All the above-described phenomena occur at normal incidence on the interface. At oblique incidence (Fig.~\ref{Fig:Concept}b), the transverse components of the wave vectors underlying the ST wave-packet are no longer invariant at the interface. Nevertheless, after an appropriate transformation a law of refraction for ST wave packets at oblique incidence can be formulated. If $\phi_{1}$ is the angle of incidence and $\phi_{2}$ is the corresponding angle in the second medium (with $n_{1}\sin{\phi_{1}}\!=\!n_{2}\sin{\phi_{2}}$), then the relationship between $\widetilde{n}_{1}$ and $\widetilde{n}_{2}$ takes the form:
\begin{equation}\label{Eq:ObliqueSnellsLawBaseband}
n_{1}(n_{1}-\widetilde{n}_{1})\cos^{2}{\phi_{1}}=n_{2}(n_{2}-\widetilde{n}_{2})\cos^{2}{\phi_{2}}.
\end{equation}
Just as for normal incidence, the subluminal-to-superluminal threshold cannot be crossed at oblique incidence. The effects discussed above hold for oblique incidence after the appropriate adjustments. For example, the group-index threshold $\widetilde{n}_{\mathrm{th}}(\phi_{1})$ is reduced with respect to $\widetilde{n}_{\mathrm{th}}(0)\!=\!n_{1}+n_{2}$ by a factor  $1+\tfrac{n_{1}}{n_{2}}\sin^{2}{\phi_{1}}\!>\!1$. However, a new phenomenon emerges at oblique incidence: $\widetilde{n}_{2}$ depends on $\phi_{1}$. In other words, the group velocity in the second medium $\widetilde{v}_{2}$ now varies with the angle of incidence $\phi_{1}$ in the first medium, even when $\widetilde{v}_{1}$ is held fixed. When $n_{1}\!<\!n_{2}$, $\widetilde{n}_{2}(\phi_{1})$ increases with $\phi_{1}$ in the superluminal regime, and decreases with $\phi_{1}$ in the subluminal regime (the opposite trends occur when $n_{1}\!>\!n_{2}$). We verify these predictions in Fig.~\ref{Fig:ObliqueIncidence}a where we plot $\Delta\widetilde{n}_{2}\!=\!\widetilde{n}_{2}(\phi_{1})-\widetilde{n}_{2}(0)$ for subluminal and superluminal wave packets obliquely incident from free space to sapphire.


The change in $\widetilde{v}_{2}$ with $\phi_{1}$ leads to a remarkable consequence related to the optical synchronization of multiple remote receivers. The envisioned scenario is depicted in Fig.~\ref{Fig:ObliqueIncidence}b-c, where a transmitting station at a distance $d_{1}$ from the interface (with $n_{1}\!<\!n_{2}$) sends a pulse at different incidence angles to reach receiving stations at different positions at a fixed depth $d_{2}$ beyond the interface. Can the pulse reach the receivers simultaneously? This is of course impossible when using traditional pulses: the distances are different whereas the group velocities are fixed. Surprisingly, the law of refraction in Eq.~\ref{Eq:ObliqueSnellsLawBaseband} enables fulfilling this task. If the group-delay difference between two paths in the first medium is $\Delta\tau_{1}$ and in the second medium $\Delta\tau_{2}$, then synchronizing the receivers requires that $\Delta\tau_{1}+\Delta\tau_{2}\!=\!0$. That is, the extra delay in the longer path in the first medium must be compensated by a reduced delay in the second, which requires that $\widetilde{v}_{2}$ increase with $\phi_{1}$. This latter requirement is satisfied in the subluminal regime as verified experimentally in Fig.~\ref{Fig:ObliqueIncidence}a. We plot in Fig.~\ref{Fig:ObliqueIncidence}d the sum $\Delta\tau\!=\!\Delta\tau_{1}+\Delta\tau_{2}$ while varying $\widetilde{n}_{1}$ and $\phi_{1}$. Realizing $\Delta\tau\!\approx\!0$ is possible over a wide range of incident angles $\phi_{1}$ for a specific $\widetilde{n}_{1}$, signifying that the wave packet reaches simultaneously all such receivers at the selected depth.

Our findings apply to ST wave packets independently of the details of their external degrees of freedom, which lends support to considering ST wave packets as objects in their own right identified by an internal degree of freedom, namely the spectral tilt angle. The rich physics of refraction of ST wave packets hints at exciting possibilities in remote sensing, subsurface imaging, optical synchronization, synthetic aperture radars, and phased-array radars, which is made all the more possible by the recent realization of extended propagation distances (reaching $\sim\!70$~m \cite{Bhaduri19OL}) and large differential group delays (a delay-bandwidth product of $\sim\!100$ \cite{Yessenov19OE}). With the law of refraction for ST wave packets established, it can be exploited in designing optical devices tailored for harnessing the unique features of such fields, exploring new vistas for controlling light-matter interactions, and examining the propagation of ST wave packets in graded-index materials, epsilon-near-zero materials \cite{Liberal17NP}, and metasurfaces \cite{Yu11Science}. Finally, we have couched our work here in terms of optical waves, but these results are equally applicable to other wave phenomena, such as acoustics, ultrasonics \cite{Parker16OE}, and even quantum-mechanical wave functions.

\clearpage
\bibliography{diffraction}

\vspace{0.5cm}
\noindent
\textbf{Acknowledgments.} We thank Demetrios N. Christidoulides, Aristide Dogariu, and Kenneth L. Schepler for useful discussions. This work was supported by the U.S. Office of Naval Research (ONR) under contract N00014-17-1-2458.

\vspace{0.5cm}
\noindent
The Supplementary Information provides the theoretical background and derivations, the detailed experimental setup and procedure, and further experimental results.

\clearpage

\clearpage

\clearpage

\clearpage

\clearpage

\begin{center}
   \large \textbf{Anomalous refraction of optical space-time wave packets:\\Supplementary Material}
\end{center}

\begin{center}
    Basanta Bhaduri, Murat Yessenov, and Ayman F. Abouraddy \\
   \textit{CREOL, The College of Optics \& Photonics,\\ University of Central Florida, Orlando, Florida 32816, USA}
\end{center}


\renewcommand{\thepage}{S\arabic{page}}  
\renewcommand{\thesection}{S\arabic{section}}   
\renewcommand{\thetable}{S\arabic{table}}   
\renewcommand{\thefigure}{S\arabic{figure}}
\renewcommand{\theequation}{S\arabic{equation}}  

\setcounter{equation}{0}    
\setcounter{figure}{0}    
\setcounter{page}{1}

\clearpage

\tableofcontents

\section{Plane-wave decomposition of space-time wave packets}

\subsection{ST wave packets in free space}

Consider a generic scalar pulsed optical beam described by the electric field $E(x,z,t)$, where $x$ and $z$ are the transverse and longitudinal coordinates, and $t$ is time. The field is assumed to be uniform along the other transverse dimension $y$, which is henceforth dropped from all expressions for the field. We write the field in terms of a carrier and a slowly varying envelope $E(x,z,t)\!=\!e^{i(k_{\mathrm{o}}z-\omega_{\mathrm{o}}t)}\psi(x,z,t)$, where $\omega_{\mathrm{o}}$ is a fixed carrier frequency, $k_{\mathrm{o}}\!=\!\omega_{\mathrm{o}}/c$ is a fixed wave number, and $c$ is the speed of light in vacuum. The envelope can be decomposed into plane waves as follows:
\begin{equation}
\psi(x,z,t)=\!\iint\!dk_{x}d\Omega\,\widetilde{\psi}(k_{x},\Omega)\,e^{i(k_{x}x+[k_{z}-k_{\mathrm{o}}]z-\Omega t)},
\end{equation}
where $\widetilde{\psi}(k_{x},\Omega)$ is the spatio-temporal Fourier transform of $\psi(x,0,t)$, $k_{x}$ is the transverse component of the wave vector (referred to as the spatial frequency), $k_{z}$ is the axial component of the wave number, $\Omega\!=\!\omega-\omega_{\mathrm{o}}$ is the temporal frequency with respect to the carrier, and $\omega$ is the temporal frequency. The components of the wave vector satisfy the free-space dispersion relationship $k_{x}^{2}+k_{z}^{2}\!=\!(\omega/c)^{2}$, which corresponds geometrically to the surface of a cone that we refer to as the `light-cone'; Fig.~\ref{Fig:SpectralHyperplanesInFreeSpaceAndInMaterial}a. The spatio-temporal spectrum of the field $\widetilde{E}(k_{x},\omega)\!=\!\widetilde{E}(k_{x},\Omega+\omega_{\mathrm{o}})\!=\!\widetilde{\psi}(k_{x},\Omega)$ can be represented in general by a two-dimensional domain on the surface of the light-cone.

Two instructive cases can be examined in which the dimensionality of the 2D spatio-temporal spectrum is reduced to one. First, \textit{monochromatic} beams $\widetilde{\psi}(k_{x},\Omega)\rightarrow\widetilde{\psi}(k_{x})\delta(\Omega)$ lie at the intersection of the light-cone with the horizontal iso-frequency plane $\omega\!=\!\omega_{\mathrm{o}}$. Such a beam has a spatial bandwidth $\Delta k_{x}$, but no temporal bandwidth. Second, \textit{pulsed plane waves} $\widetilde{\psi}(k_{x},\Omega)\rightarrow\widetilde{\psi}(\Omega)\delta(k_{x})$ lie along the light-line $\omega/c\!=\!k_{z}$, which is the tangent of the light-cone at $k_{x}\!=\!0$. Such a pulse has a temporal bandwidth $\Delta\Omega$, but no spatial bandwidth. In both cases, the two-dimensional spatio-temporal spectrum is reduced either to a purely spatial spectrum (monochromatic beams) or a purely temporal spectrum (pulsed plane wave).

The ST wave packets we investigate in the main text also have a reduced dimensionality when compared to traditional pulsed beams; however, they retain finite spatial \textit{and} temporal bandwidths. The spatio-temporal spectrum of ST wave packets lies along the intersection of the light-cone with the tilted spectral hyperplane $\mathcal{P}(\theta)$ given by the equation
\begin{equation}
\frac{\omega}{c}\!=\!k_{\mathrm{o}}+(k_{z}-k_{\mathrm{o}})\tan{\theta},
\end{equation}
corresponding to a plane that is parallel to the $k_{x}$-axis and $\theta$ is the angle with the $k_{z}$-axis \cite{Donnelly93PRSLA,Efremidis17OL,PorrasPRA18}. The plane intersects with the light-cone at the point $(k_{x},k_{z},\omega/c)\!=\!(0,k_{\mathrm{o}},k_{\mathrm{o}})$. The result of this intersection is a conic section that depends on the spectral tilt angle $\theta$; Fig.~\ref{Fig:SpectralHyperplanesInFreeSpaceAndInMaterial}b. Furthermore, the group velocity $\widetilde{v}\!=\!\tfrac{\partial\omega}{\partial k_{z}}$ \cite{SaariPRA18} and the group index $\widetilde{n}\!=\!c/\widetilde{v}$ of the ST wave packet are determined by $\theta$: $\widetilde{v}\!=\!c\tan{\theta}$ and $\widetilde{n}\!=\!\cot{\theta}$. The plane-wave expansion of the wave packet envelope is
\begin{equation}
\psi(x,z,t)\!=\!\int\!dk_{x}\,\widetilde{\psi}(k_{x})\,e^{ik_{x}x}e^{-i\Omega(t-\frac{z}{c}\cot{\theta})}\!=\!\psi(x,0,t-z/\widetilde{v}),
\end{equation}
which corresponds to a wave packet transported rigidly along the $z$ direction at a group velocity $\widetilde{v}$ without diffraction or dispersion \cite{Longhi04OE,Saari04PRE,Turunen10PO,FigueroaBook14}.

\subsection{Families of ST wave packets}

\begin{figure*}[t!]
\centering
\includegraphics[width=9.6cm]{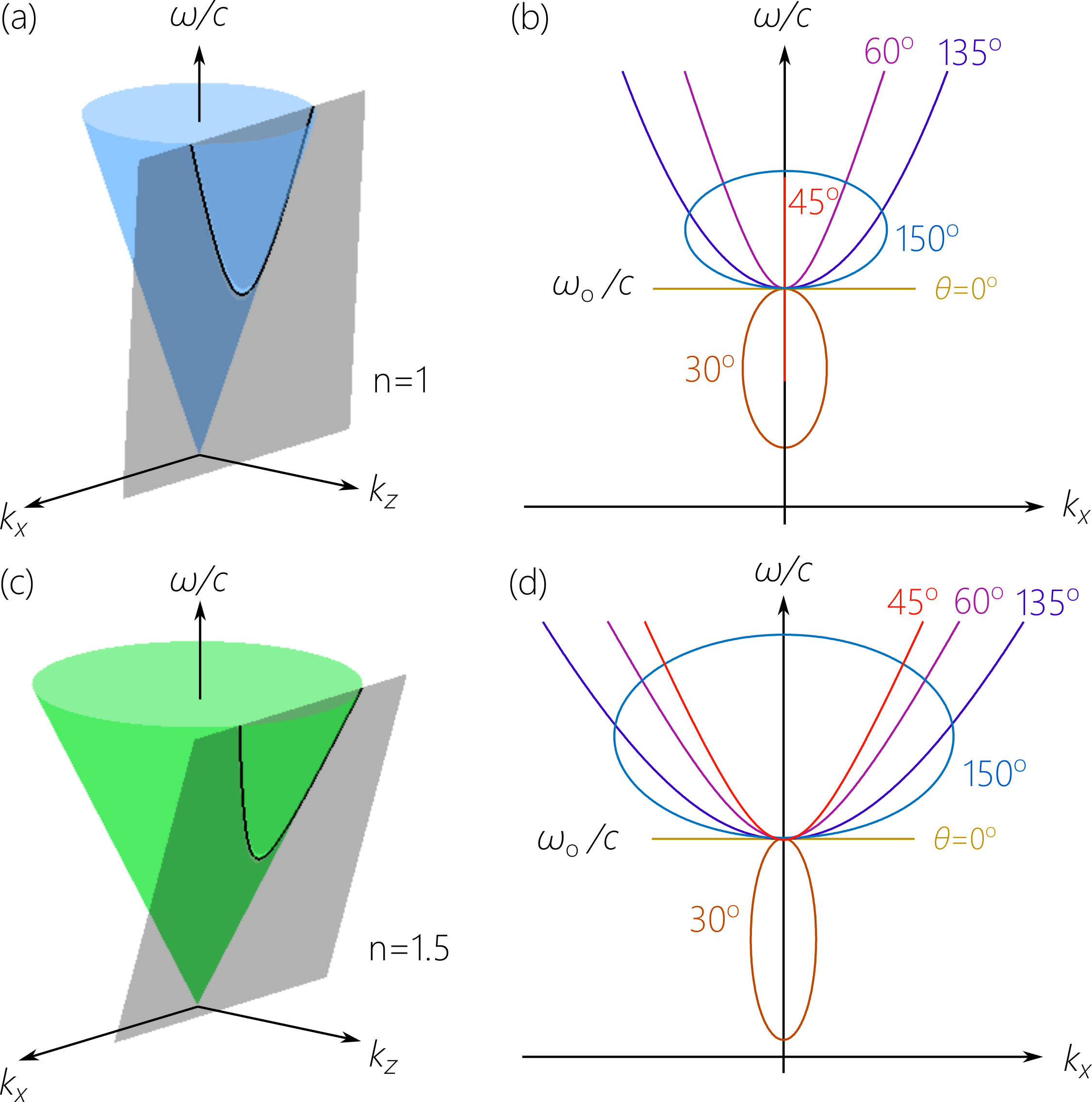}
\caption{Concept of ST wave packets. (a) The spectral hyperplane $\mathcal{P}(\theta)$ intersects with the free-space light-cone $k_{x}^{2}+k_{z}^{2}\!=\!(\tfrac{\omega}{c})^{2}$. (b) Projections of the spatio-temporal spectral trajectories associated with ST wave packets onto the $(k_{x},\tfrac{\omega}{c})$-plane for different $\theta$. Our experimental spatio-temporal synthesis procedure implements these correlation functions between spatial and temporal frequencies, $k_{x}$ and $\omega$, respectively. (c) The spectral hyperplane $\mathcal{P}(\theta)$ intersects with the light-cone in a material of index $n$, $k_{x}^{2}+k_{z}^{2}\!=\!n^{2}(\tfrac{\omega}{c})^{2}$; here $n\!=\!1.5$. (d) Same as (b) for the material light-cone in (c).}
\label{Fig:SpectralHyperplanesInFreeSpaceAndInMaterial}
\end{figure*}

The spectral hyperplane $\mathcal{P}$ generates a family of ST wave packets that we refer to as `baseband' because the spatial frequency $k_{x}\!=\!0$ and its vicinity of low spatial frequencies are physically allowed to contribute to the spatial spectrum of the wave packet. The type of conic section at the intersection of the light-cone with $\mathcal{P}(\theta)$ depends on $\theta$: it is a circle corresponding to monochromatic beams when $\theta\!=\!0$; an ellipse corresponding to positive subluminal ST wave packets when $0\!<\!\theta\!<\!45^{\circ}$; a tangential line corresponding to luminal pulsed plane wave when $\theta\!=\!45^{\circ}$; a hyperbola when $45^{\circ}\!<\!\theta\!<\!135^{\circ}$, with positive superluminal group velocity when $45^{\circ}\!<\!\theta\!<\!90^{\circ}$ and negative superluminal group velocity when $90^{\circ}\!<\!\theta\!<\!135^{\circ}$, separated by the singularity at $\theta\!=\!90^{\circ}$ where the group index is formally infinite; a parabola corresponding to negative luminal ST wave packets at $\theta\!=\!135^{\circ}$; and an ellipse corresponding to negative subluminal ST wave packets when $135^{\circ}\!<\!\theta\!<\!180^{\circ}$; see Fig.~\ref{Fig:SpectralHyperplanesInFreeSpaceAndInMaterial}b. In general, the equation for the conic section projected onto the $(k_{x},\omega/c)$-plane is
\begin{equation}
\frac{(1+\tan{\theta})^{2}}{k_{\mathrm{o}}^{2}\tan^{2}{\theta}}\left(\frac{\omega}{c}-\frac{k_{\mathrm{o}}}{1+\tan{\theta}}\right)^{2}+\frac{1+\tan{\theta}}{1-\tan{\theta}}\,\,\,\,\frac{k_{x}^{2}}{k_{\mathrm{o}}^{2}}=1.
\end{equation}
See Ref.~\cite{Kondakci16OE,Kondakci17NP,Kondakci19NC,Yessenov19PRA} for details.

Two other distinct families of ST wave packets are possible: X-waves that are produced by the spectral hyperplane given by the equation $\omega/c\!=k_{z}\tan{\theta}$, which passes through the origin and intersects with the light-cone in a pair of lines meeting at the origin \cite{Lu92IEEEa,Turunen10PO}; and `sideband' ST wave packets generated by the hyperplane $\omega/c\!=\!k_{\mathrm{o}}+(k_{z}+k_{\mathrm{o}})\tan{\theta}$, which passes through the point $(k_{x},k_{z},\omega/c)\!=\!(0,-k_{\mathrm{o}},k_{\mathrm{o}})$, whereupon $k_{x}\!=\!0$ and its vicinity of low spatial frequencies are physically forbidden from contributing to the spatial spectrum of the ST wave packet ($k_{z}\!<\!0$ is incompatible with causal excitation and propagation) \cite{Brittingham83JAP,Turunen10PO,Yessenov19PRA}. The group velocity and group index are still related to $\theta$ through $\widetilde{v}\!=\!c\tan{\theta}$ and $\widetilde{n}\!=\!\cot{\theta}$, respectively. In both cases of X-waves and sideband ST wave packets, $\theta$ is confined to the range $45^{\circ}\!<\!\theta\!<\!90^{\circ}$, thus corresponding only to positive superluminal group velocities \cite{Yessenov19PRA}. We thus eschewed examining these two classes in favor of baseband ST wave packets, and will investigate them in detail elsewhere.

\subsection{ST wave packets in a non-dispersive material}

\begin{figure*}[bh]
\centering
\includegraphics[width=12.6cm]{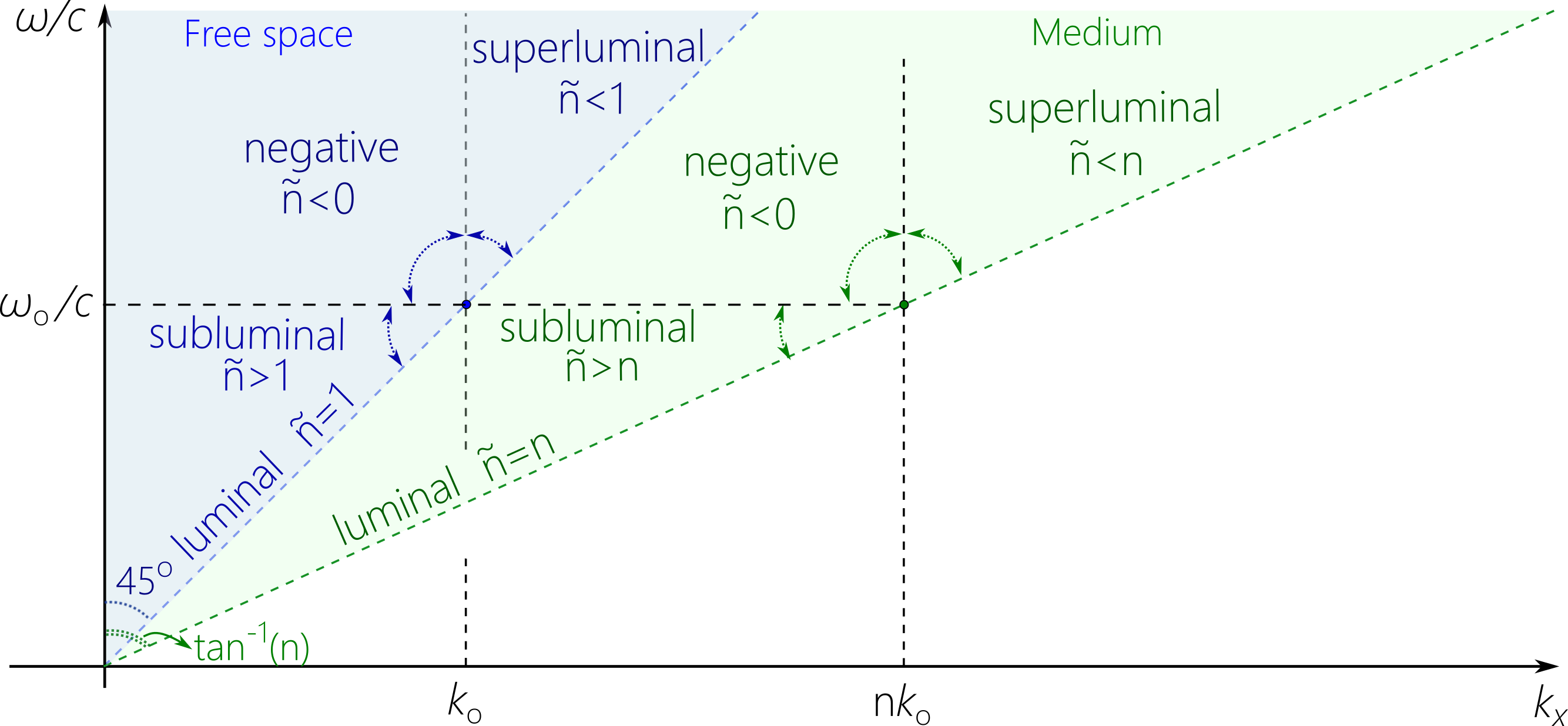}
\caption{Domains of subluminal, luminal, superluminal, and negative-$\widetilde{v}$ ST wave packets in free space and in a material of refractive index $n$, all represented in the $(k_{z},\tfrac{\omega}{c})$-plane. The projection of the free-space light-cone is to the left (blue shading) and the that of the material light-cone is to the right (green shading). The subluminal domain corresponds to the angular range where $\widetilde{n}\!<\!n$, angles between the horizontal axis and the light line; luminal to $\widetilde{n}\!=\!n$; superluminal to the range between the light line and the vertical axis $\widetilde{n}\!>\!n$; and negative-$v_{\mathrm{g}}$ to the range from the vertical axis counter-clockwise to the horizontal axis $\widetilde{n}\!<\!0$.)}
\label{Fig:ProjectionsOntoThePlanesInMaterial}
\end{figure*}

In a material of refractive index $n$, the light-cone becomes $k_{x}^{2}+k_{z}^{2}\!=\!n^{2}(\omega/c)^{2}$, and the plane $\mathcal{P}(\theta)$ associated with a ST wave packet is given by the equation
\begin{equation}
\omega/c\!=\!k_{\mathrm{o}}+(k_{z}-nk_{\mathrm{o}})\tan{\theta}.
\end{equation}
This plane passes through the point $(k_{x},k_{z},\tfrac{\omega}{c})\!=\!(0,nk_{\mathrm{o}},k_{\mathrm{o}})$, which lies on the light-line in the medium; Fig.~\ref{Fig:SpectralHyperplanesInFreeSpaceAndInMaterial}c. Critically, the expressions for the group velocity and group index are not changed and depend solely on $\theta$ independently of $n$. The ranges for $\theta$ that dictate the nature of the conic section at the intersection of the light-cone with $\mathcal{P}(\theta)$ are modified: an ellipse corresponding to positive subluminal ST wave packets when $0\!<\!\theta\!<\!\arctan{\tfrac{1}{n}}$; a tangential line corresponding to luminal pulsed plane wave when $\theta\!=\!\arctan{\tfrac{1}{n}}$; a hyperbola when $\arctan{\tfrac{1}{n}}\!<\!\theta\!<\!180^{\circ}-\arctan{\tfrac{1}{n}}$, with positive superluminal group velocity when $\arctan{\tfrac{1}{n}}\!<\!\theta\!<\!90^{\circ}$ and negative superluminal group velocity when $90^{\circ}\!<\!\theta\!<\!180^{\circ}-\arctan{\tfrac{1}{n}}$, separated by a singularity at $\theta\!=\!90^{\circ}$ where the group index is formally infinite; a parabola corresponding to negative luminal ST wave packets at $\theta\!=\!180^{\circ}-\arctan{\tfrac{1}{n}}$; and an ellipse corresponding to negative subluminal ST wave packets when $180^{\circ}-\arctan{\tfrac{1}{n}}\!<\!\theta\!<\!180^{\circ}$; Fig.~\ref{Fig:SpectralHyperplanesInFreeSpaceAndInMaterial}d. Also, the carrier phase term for the field becomes $e^{i(nk_{\mathrm{o}}z-\omega_{\mathrm{o}}t)}$. The general equation for the conic section at the intersection of the light-cone of the material and the plane $\mathcal{P}(\theta)$ is given by \cite{Bhaduri19Optica}
\begin{equation}\label{Eq:ConicSectionInMaterial}
\frac{(1+n\tan{\theta})^{2}}{n^{2}k_{\mathrm{o}}^{2}\tan^{2}{\theta}}\left(\frac{\omega}{c}-\frac{k_{\mathrm{o}}}{1+n\tan{\theta}}\right)^{2}+\frac{1+n\tan{\theta}}{1-n\tan{\theta}}\,\,\,\,\frac{k_{x}^{2}}{n^{2}k_{\mathrm{o}}^{2}}=1.    
\end{equation}

\subsection{Interpretation of the projections of the spatio-temporal spectrum}

It is critical to appreciate the interpretation of the projections of the spatio-temporal spectrum of a ST wave packet onto the $(k_{z},\omega/c)$-plane, specifically with change in the refractive index $n$. The projection onto the $(k_{z},\omega/c)$-plane determines the group velocity $\widetilde{v}$, whereas the projection onto the $(k_{x},\omega/c)$-plane is that used in synthesizing the wave packet and remains invariant upon normal incidence on a planar interface. We have plotted in Fig.~\ref{Fig:ProjectionsOntoThePlanesInMaterial} the various domains for ST wave packets (subluminal, luminal, superluminal, and negative-$\widetilde{v}$) in free space and in a material of refractive index $n$ projected onto the $(k_{z},\omega/c)$-plane.

\section{Refraction of space-time wave packets at normal incidence}

As described above, the spatio-temporal spectral trajectory for any baseband ST wave packets lies at the intersection of the light-cone with a tilted spectral hyperplane $\mathcal{P}(\theta)$, where $\mathcal{P}$ is parallel to the $k_{x}$-axis and $\theta$ is the spectral tilt angle with respect to the $k_{z}$-axis. In a medium of refractive index $n$, the projection of this intersection onto the $(k_{x},\tfrac{\omega}{c})$-plane is the conic section given in Eq.~\ref{Eq:ConicSectionInMaterial}, which applies for all $\theta$.

\begin{figure*}[t!]
\centering
\includegraphics[width=17.6cm]{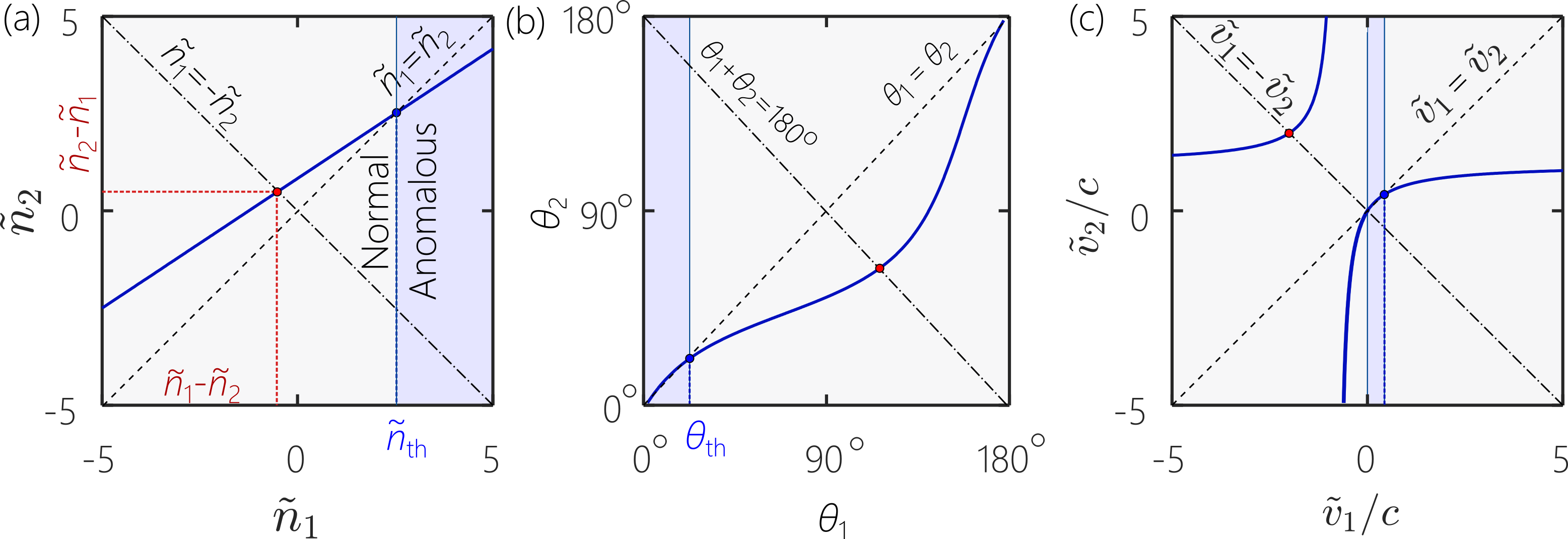}
\caption{(a) Plot of the law of refraction for ST wave packets in terms of the group indices $\widetilde{n}_{1}$ and $\widetilde{n}_{2}$; (b) in terms of the spectral tilt angles $\theta_{1}$ and $\theta_{2}$; and (c) in terms of the group velocities $\widetilde{v}_{1}$ and $\widetilde{v}_{2}$. The features are generic, but for concreteness we consider $n_{1}\!<\!n_{2}$. The corresponding plots for $n_{1}\!>\!n_{2}$ can be obtained by flipping the curves around the diagonal.}
\label{Fig:LawOfRefractionInDifferentVariables}
\end{figure*}

In our work, we consider ST wave packets with small spatial bandwidths $\Delta k_{x}\!\ll\!k_{\mathrm{o}}$. Because of the close connection between spatial and temporal frequencies, this also implies a small temporal bandwidth, $\Delta\omega\!\ll\!\omega_{\mathrm{o}}$. We can thus approximate the conic section given in Eq.~\ref{Eq:ConicSectionInMaterial} in the vicinity of $k_{x}\!=\!0$ by a parabola,
\begin{equation}\label{eq:approximation}
\frac{\omega}{\omega_{\mathrm{o}}}-1=\frac{\Omega}{\omega_{\mathrm{o}}}\approx\frac{k_{x}^{2}}{2n^{2}k_{\mathrm{o}}^{2}}\,\frac{n\tan{\theta}}{n\tan{\theta}-1}=\frac{k_{x}^{2}}{2k_{\mathrm{o}}^{2}}\,\frac{1}{n(n-\widetilde{n})}.
\end{equation}
At \textit{normal} incidence at a planar interface between two materials the spatial frequencies $k_{x}$ are invariant, in addition to the invariance of the temporal frequencies $\omega$. Therefore, the quantity
$n(n-\widetilde{n})$ is in turn an invariant, and we thus have the law of refraction for baseband ST wave packets 
\begin{equation}
n_{1}(n_{1}-\widetilde{n}_{1})=n_{2}(n_{2}-\widetilde{n}_{2}).
\end{equation}
This law can be rewritten in terms of the spectral tilt angles $\theta_{1}$ and $\theta_{2}$ as follows:
\begin{equation}\label{Eq:NormalBasebandWithAngles}
n_{1}\left(n_{1}-\cot{\theta_{1}}\right)=n_{2}\left(n_{2}-\cot{\theta_{2}}\right),
\end{equation}
or in terms of the group velocities $\widetilde{v}_{1}\!=\!c\tan{\theta_{1}}$ and $\widetilde{v}_{2}\!=\!c\tan{\theta_{2}}$ as follows:
\begin{equation}
\frac{n_{2}}{\widetilde{v}_{2}}=\frac{n_{1}}{\widetilde{v}_{1}}+\frac{n_{2}^{2}-n_{1}^{2}}{c}.
\end{equation}
Figure~\ref{Fig:LawOfRefractionInDifferentVariables} depicts this law of refraction as a relationship between $(\widetilde{n}_{1},\widetilde{n}_{2})$, $(\theta_{1},\theta_{2})$, and $(\widetilde{v}_{1},\widetilde{v}_{2})$.

The approximation in Eq.~\ref{eq:approximation} only requires that the spatial bandwidth be smaller than the central wave number $\Delta k_{x}\!\ll\!k_{\mathrm{o}}$ and that the temporal bandwidth be smaller than the central frequency $\Delta\omega\!\ll\!\omega_{\mathrm{o}}$. These conditions are well satisfied in our work.

\subsection{Subluminal regime: $\widetilde{n}_{1}\!>\!n_{1}$ and $\widetilde{n}_{2}\!>\!n_{2}$}

It can be verified by direct substitution that $\widetilde{n}_{1}\!=\!\widetilde{n}_{2}\!=\!n_{1}+n_{2}\!=\!\widetilde{n}_{\mathrm{th}}$ satisfies the law of refraction given above. Furthermore, it can be shown that this special case is the only solution that produces $\widetilde{n}_{1}\!=\!\widetilde{n}_{2}$, which separates the `normal' and `anomalous' regimes. In the anomalous regime $\widetilde{n}_{1}\!>\!\widetilde{n}_{\mathrm{th}}$, we set $\widetilde{n}_{1}\!=\!\widetilde{n}_{\mathrm{th}}+\delta\widetilde{n}_{1}$ and $\widetilde{n}_{2}\!=\!\widetilde{n}_{\mathrm{th}}+\delta\widetilde{n}_{2}$, substitute in the law of refraction and obtain $\delta\widetilde{n}_{2}\!=\!\tfrac{n_{1}}{n_{2}}\delta\widetilde{n}_{1}\!<\!\delta\widetilde{n}_{1}$, such that $\widetilde{n}_{2}\!<\!\widetilde{n}_{1}$ and thus $\widetilde{v}_{2}\!>\!\widetilde{v}_{1}$; i.e., the group velocity in the second medium (of higher refractive index) \textit{increases} against traditional expectations. In the normal regime $\widetilde{n}_{1}\!<\!\widetilde{n}_{\mathrm{th}}$, we set $\widetilde{n}_{1}\!=\!\widetilde{n}_{\mathrm{th}}-\delta\widetilde{n}_{1}$ and $\widetilde{n}_{2}\!=\!\widetilde{n}_{\mathrm{th}}-\delta\widetilde{n}_{2}$, substitute in the law of refraction and obtain again $\delta\widetilde{n}_{2}\!=\!\tfrac{n_{1}}{n_{2}}\delta\widetilde{n}_{1}\!<\!\delta\widetilde{n}_{1}$, such that $\widetilde{n}_{2}\!>\!\widetilde{n}_{1}$ this time, and thus $\widetilde{v}_{1}\!<\!\widetilde{v}_{2}$; i.e., the group velocity in the second higher-index medium decreases as traditionally expected.

\subsection{Superluminal regime: $\widetilde{n}_{1}\!<\!n_{1}$ and $\widetilde{n}_{2}\!<\!n_{2}$}

\begin{figure*}[t!]
\centering
\includegraphics[width=15.6cm]{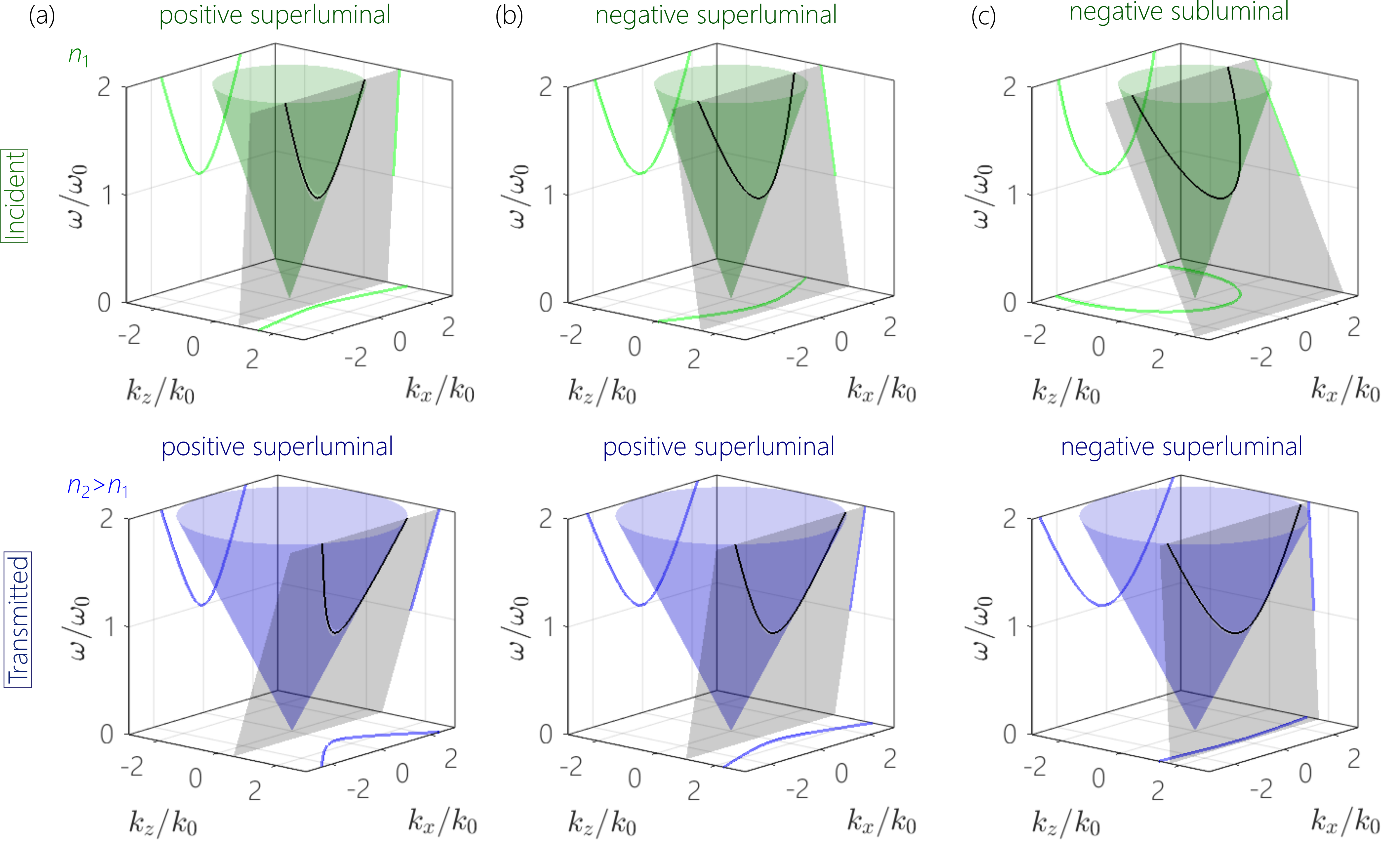}
\caption{Dynamics of the spatio-temporal spectrum of a ST wave packet as it propagates from one medium to another in different regimes of the group velocity. (a) Both incident and transmitted ST wave-packets are positive superluminal, $0\!<\!\widetilde{n}_{1}\!<\!n_{1}$ and $0\!<\widetilde{n}_{2}\!<\!n_{2}$; (b) the incident ST wave packet is negative superluminal $\widetilde{n}_{1}\!<\!0$ ($|\widetilde{n}_{1}|\!<\!n_{1}$), which is transformed into a positive superluminal transmitted ST wave packet $0\!<\!\widetilde{n}_{2}\!<\!n_{2}$; and (c) an incident negative subluminal ST wave-packet $|\widetilde{n}_{1}|\!>\!n_{1}$ is converted into a negative superluminal transmitted ST wave packet $|\widetilde{n}_{2}|\!<\!n_{2}$. The plots are obtained for $n_{1}\!=\!1$ and $n_{2}\!=\!1.5$.}
\label{Fig:LawOfRefractionSuperluminal}
\end{figure*}

We examine the case where $n_{1}\!<\!n_{2}$. Consider first the regime $0\!<\!\widetilde{n}_{1}\!<\!n_{1}$; that is, when the incident wave packet has a positive superluminal group velocity. We set $\widetilde{n}_{1}\!=\!n_{1}-\delta\widetilde{n}_{1}$ (where $0\!<\!\delta\widetilde{n}_{1}\!<\!n_{1}$) and $\widetilde{n}_{2}\!=\!n_{2}-\delta\widetilde{n}_{2}$. Substituting into the law of refraction we obtain $\delta\widetilde{n}_{2}\!=\!\tfrac{n_{1}}{n_{2}}\delta\widetilde{n}_{1}\!<\!\delta\widetilde{n}_{1}$, such that $\widetilde{n}_{2}\!>\!\widetilde{n}_{1}$ and thus $\widetilde{v}_{2}\!<\!\widetilde{v}_{1}$; that is, the wave packet slows down as expected, and both are in the positive superluminal regime. When $\widetilde{n}_{1}\!=\!0$, the transmitted wave packet has $\widetilde{n}_{2}\!=\!(n_{2}^{2}-n_{1}^{2})/n_{2}\!>\!0$.

Next, we examine the negative superluminal regime $-n_{1}\!<\!\widetilde{n}_{1}\!<\!0$. We set $\widetilde{n}_{1}\!=\!-n_{1}+\delta\widetilde{n}_{1}$ (where $0\!<\!\delta\widetilde{n}_{1}\!<\!n_{1}$) and $\widetilde{n}_{2}\!=\!n_{2}-\delta\widetilde{n}_{2}$, substitute into the law of refraction and obtain $\delta\widetilde{n}_{2}\!=\!\tfrac{n_{1}}{n_{2}}(2n_{1}-\delta\widetilde{n}_{1})$. From this we can understand the transition of the ST wave packet in the second medium into the negative superluminal regime. When the transmitted wave packet reaches the transition from positive to negative superluminal regimes $\widetilde{n}_{2}\!=\!0$ ($\delta\widetilde{n}_{2}\!=\!n_{2}$), the incident wave packet has $\widetilde{n}_{1}\!=\!-(n_{2}^{2}-n_{1}^{2})/n_{1}$. In the special case of $n_{2}\!=\!\sqrt{2}n_{1}$, the \textit{transmitted} wave packet reaching $\widetilde{n}_{2}\!=\!0$ is thus associated with the incident wave packet being negative luminal $\widetilde{n}_{1}=-n_{1}$. When $n_{2}\!<\!\sqrt{2}n_{1}$, the transmitted wave packet reaching the condition $\widetilde{n}_{2}\!=\!0$ is associated with the incident wave packet remaining negative superluminal $\widetilde{n}_{1}\!>\!-n_{1}$; whereas a larger index contrast $n_{2}\!>\!\sqrt{2}n_{1}$ implies that the transmitted wave packet reaching $\widetilde{n}_{2}\!=\!0$ is associated with the incident wave packet being negative subluminal $\widetilde{n}_{1}\!<\!-n_{1}$. In general, the larger the index-contrast, the larger the separation in the group indices of the incident and transmitted ST wave packets. Some of the cases discussed above are depicted in Fig.~\ref{Fig:LawOfRefractionSuperluminal}.

\section{Refraction of space-time wave packets at oblique incidence}

\begin{figure*}[t!]
\centering
\includegraphics[width=12.6cm]{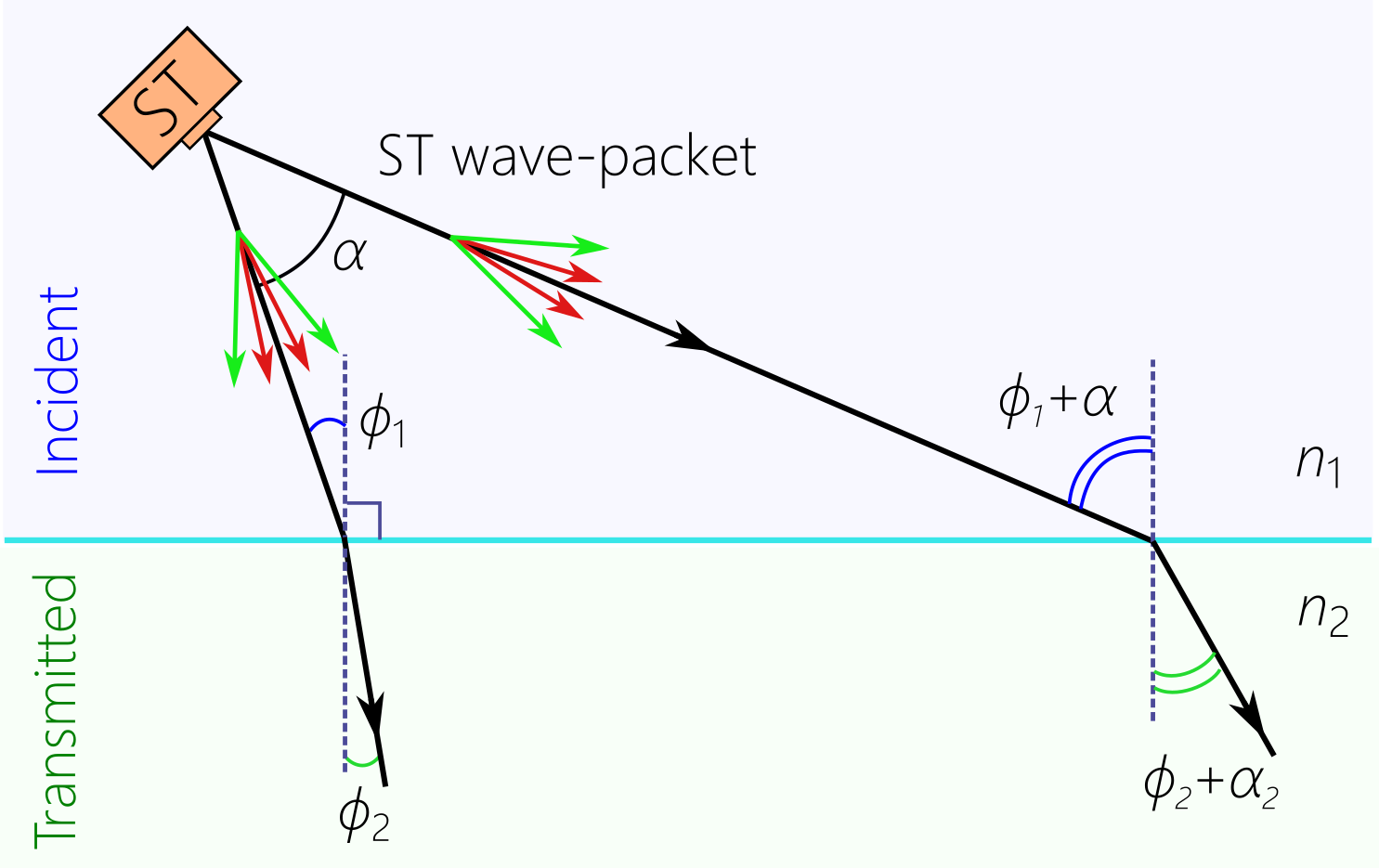}
\caption{Illustration of the geometry of the oblique-incidence configuration.}
\label{Fig:GeometryOfObliqueIncidence}
\end{figure*}

The geometry of the problem is illustrated in Fig.~\ref{Fig:GeometryOfObliqueIncidence}. Although the parabolic approximation of the spatio-temporal spectra of the ST wave packet with respect to the propagation axis in both materials still holds and the frequencies $\omega$ are invariant across the interface, we no longer have the spatial frequencies $k_{x}$ of the ST wave packet invariant with respect to the interface. The invariance of the frequencies $\omega$ implies that
\begin{equation}\label{Eq:InvarianceOfOmega}
\frac{k_{x1}^{2}}{n_{1}(n_{1}-\widetilde{n}_{1})}=\frac{k_{x2}^{2}}{n_{2}(n_{2}-\widetilde{n}_{2})}; 
\end{equation}
however, $k_{x1}\!\neq\!k_{x2}$, where $k_{x1}$ and $k_{x2}$ are considered with respect to the propagation axis of each ST wave packet separately. Instead, we need to first change the coordinate system to that measured with respect to the normal to the interface.

The angles made by the monochromatic plane-wave components in each ST wave packet with respect to its own propagation axis are labeled $\alpha$: $k_{x1}\!=\!n_{1}k_{\mathrm{o}}\sin{\alpha_{1}}$ and $k_{x2}\!=\!n_{2}k_{\mathrm{o}}\sin{\alpha_{2}}$. The parabolic approximation is equivalent to a first-order Taylor series of the sine functions, such that $k_{x1}\!\approx\!n_{1}k_{\mathrm{o}}\alpha_{1}$ and $k_{x2}\!\approx\!n_{2}k_{\mathrm{o}}\alpha_{2}$. The angle of incidence of the ST wave packet in the first material $\phi_{1}$ is that of its propagation axis, and similarly for $\phi_{2}$ in the second material. Snell's law implies that $n_{1}\sin{\phi_{1}}\!=\!n_{2}\sin{\phi_{2}}$. We \textit{do not make any small-angle approximation} with respect to $\phi_{1}$ and $\phi_{2}$.

Snell's law at the interface between the two materials for each constitutive monochromatic plane wave is
\begin{equation}
n_{1}\sin{(\phi_{1}+\alpha_{1})}=n_{2}\sin{(\phi_{2}+\alpha_{2})}.
\end{equation}
By first expanding the sine functions $\sin{(\phi_{1}+\alpha_{1})}\!=\!\sin{\phi_{1}}\cos{\alpha_{1}}+\cos{\phi_{1}}\sin{\alpha_{1}}$ and similarly for $\sin{(\phi_{2}+\alpha_{2})}$, and then using the approximations $\sin{\alpha}\!\approx\!\alpha$ and $\cos{\alpha_{2}}\!\approx\!1$, we obtain
\begin{equation}
n_{1}\alpha_{1}\cos{\phi_{1}}\!=\!n_{2}\alpha_{2}\cos{\theta_{2}},   
\end{equation}
such that $k_{x2}=n_{1}k_{\mathrm{o}}\alpha_{1}\cos{\phi_{1}}/\cos{\phi_{2}}=k_{x1}\cos{\phi_{1}}/\cos{\phi_{2}}$. Substitution into Eq.~\ref{Eq:InvarianceOfOmega} yields the law of refraction for ST wave packets at oblique incidence,
\begin{equation}
n_{1}(n_{1}-\widetilde{n}_{1})\cos^{2}{\phi_{1}}=n_{2}(n_{2}-\widetilde{n}_{2})\cos^{2}{\phi_{2}}.
\end{equation}
The squared-cosine factors therefore result from the parabolic relationship between spatial and temporal frequencies.

In terms of the spectral tilt angles, this law of refraction is 
\begin{equation}
n_{1}\left(n_{1}-\cot{\theta_{1}}\right)\cos^{2}{\phi_{1}}=n_{2}\left(n_{2}-\cot{\theta_{2}}\right)\cos^{2}{\phi_{2}}.
\end{equation}
In terms of the group velocities, this law of refraction can be written as:
\begin{equation}
n_{1}\left(n_{1}-\frac{c}{\widetilde{v}_{1}}\right)\cos^{2}{\phi_{1}}=n_{2}\left(n_{2}-\frac{c}{\widetilde{v}_{1}}\right)\cos^{2}{\phi_{2}},
\end{equation}
which can be re-expressed as follows:
\begin{equation}
\frac{n_{2}}{\widetilde{v}_{2}}=\frac{n_{1}}{\widetilde{v}_{1}}g_{12}^{2}+\frac{n_{2}^{2}-n_{1}^{2}g_{12}^{2}}{c},
\end{equation}
where we have introduced the the parameter $g_{12}\!=\!\tfrac{\cos{\phi_{1}}}{\cos{\phi_{2}}}$. 

\section{Origin of the anomalous refraction effect at normal incidence}

In this Section we provide a description of the anomalous refraction of ST wave packets through a consideration of the dynamics of the spectral representation on the light-cone with variations in the refractive index $n$. We describe the effect from a purely \textit{geometric} standpoint. We consider a planar curve traced on the surface of a cone and examine the changes in the projection of this curve on a plane upon changing the cone apex angle if the projections of this curve are constrained in the other two planes. We find that the inflation of the cone surface with increased cone angle does \textit{not} always lead to an expansion of the projection of the curve onto that plane -- indeed, surprisingly, the projection might even shrink within certain ranges.

The question regarding the changes that ST wave packets undergo when traveling from one material to another in essence relates to the geometry of points on the surface of a cone when the apex angle changes. Assuming that the cone axis coincides with one axis of a Cartesian coordinate system, how does the distance -- projected along one axis -- between two points on the surface of the cone change with increased apex angle \textit{if we constrain their separation to be fixed when projected onto the other two axes}? Surprisingly, we find that their separation does not change monotonically with the apex angle, which results in the anomalous refraction effect described in the main text.

Consider the cone shown in Fig.~\ref{Fig:GenericConeOnTheSameRay} in a generic space $(x,y,z)$. The equation of the cone is $x^{2}+y^{2}\!=\!n^{2}z^{2}$, where the apex angle of the cone is $\tan^{-1}{n}$, the cone axis coincides with the $z$-axis, and the apex is at the origin. Increasing $n$ results in an increase in the cone angle and hence an inflation in the cone surface. Consider two points $\mathrm{P}_{1}\!=\!(x_{1},y_{1},z_{1})$ and $\mathrm{P}_{2}\!=\!(x_{2},y_{2},z_{2})$ that lie on the surface of the cone such that $x_{1}^{2}+y_{1}^{2}\!=\!n^{2}z_{1}^{2}$ and $x_{2}^{2}+y_{2}^{2}\!=\!n^{2}z_{2}^{2}$. We examine the change in the relative locations of these two points along the $y$-axis as $n$ increases under the following two constraints:
\begin{enumerate}
    \item The $z$-coordinates of $\mathrm{P}_{1}$ and $\mathrm{P}_{2}$ are held fixed ($\Delta z\!=\!z_{2}-z_{1}$ is invariant).
    \item The $x$-coordinates of $\mathrm{P}_{1}$ and $\mathrm{P}_{2}$ are held fixed ($\Delta x\!=\!x_{2}-x_{1}$ is invariant).
\end{enumerate}
We are interested in the change in the quantity $\Delta y\!=\!y_{2}-y_{1}$ with $n$. At first glance it may appear that $\Delta y$ always increases with $n$. We proceed to show that this is not necessarily the case.

\begin{enumerate}
\item
\textit{Points along a ray}. Consider the case when $\mathrm{P}_{1}$ and $\mathrm{P}_{2}$ lie along a ray as in Fig.~\ref{Fig:GenericConeOnTheSameRay}. In other words, the line connecting the two points passes through the origin. We can always rotate the coordinate system such that $x_{1}\!=\!x_{2}\!=\!0$, in which case $y_{1}\!=\!nz_{1}$, $y_{2}\!=\!nz_{2}$, and $\Delta y\!=\!n\Delta z$. In this case, $\Delta y$ always \textit{increases} with $n$ as expected. The same applies to pairs of points on \textit{any} ray, even if they do not lie in the $(y,z)$-plane.

\begin{figure*}[b!]
\centering
\includegraphics[width=9.6cm]{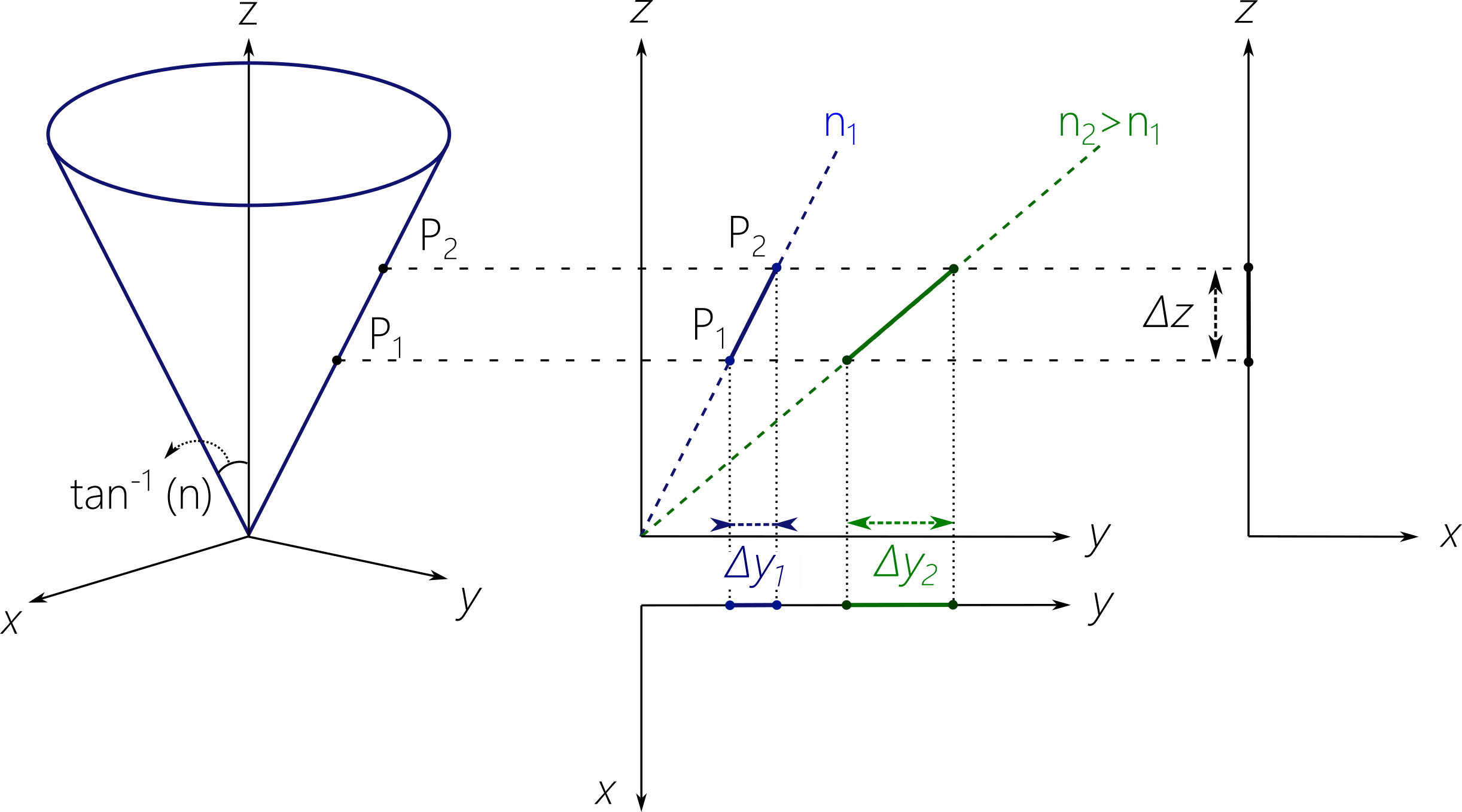}
\caption{Impact of the increase in the cone angle on two points P$_1$ and P$_2$ located on the same ray when their $x$-coordinates and $z$-coordinates are fixed.}
\label{Fig:GenericConeOnTheSameRay}
\end{figure*}

\item
\textit{Points at a fixed height}. Consider two points at the same height $z_{1}\!=\!z_{2}\!=\!z$ as in Fig.~\ref{Fig:GenericConeSameHeight}. We rotate the coordinate system such that $x_{2}\!=\!0$, resulting in
\begin{equation}
\Delta y=y_{2}-y_{1}=nz-\sqrt{n^{2}z^{2}-x_{1}^{2}}\,\,\approx\,\frac{x_{1}^{2}}{2nz}.
\end{equation}
In contrast to the previous special case, the separation $\Delta y$ \textit{decreases} with $n$. Despite the inflation of the cone with $n$, the separation between $\mathrm{P}_{1}$ and $\mathrm{P}_{2}$ projected onto the $y$-axis decreases. The reason is that the two points lie on the same circle on the cone surface at a fixed height. This circle \textit{increases} in diameter with $n$, but the projected separation is related to the \textit{curvature} of the circle, which does \textit{decrease} with $n$.

\item
\textit{General point locations}. We now consider two general locations for $\mathrm{P}_{1}$ and $\mathrm{P}_{2}$ (Fig.~\ref{Fig:GenericConeGeneralPoints}), but rotate the coordinate system -- without loss of generality -- such that $x_{2}\!=\!0$, resulting in 
\begin{equation}
\Delta y=y_{2}-y_{1}=nz_{2}-\sqrt{n^{2}z_{1}^{2}-x_{1}^{2}}\approx n\Delta z+\frac{x_{1}^{2}}{2nz_{1}}.
\end{equation}
The separation between $\mathrm{P}_{1}$ and $\mathrm{P}_{2}$ projected onto the $y$-axis is the sum of two terms, one that increases with $n$ and is a result of the inflation of the cone surface that increases the projected separation between points on a ray, and a second term that decreases with $n$ as a results of the initial separation between the two points along the $x$-axis. Because $\Delta z$, $z_{1}$, and $x_{1}$ are invariant, we can rewrite this last equation as
\begin{equation}
\frac{\Delta y}{\Delta z}=\widetilde{n}=n+\frac{g}{n},
\end{equation}
where $g\!=\!\tfrac{x_{1}^{2}}{2z_{1}\Delta z}$ is an invariant quantity independent of $n$. Taking two cones with angles $\tan^{-1}{n_{1}}$ and $\tan^{-1}{n_{2}}$, we have $\widetilde{n}_{1}\!=\!n_{1}+g/n_{1}$ and $\widetilde{n}_{2}\!=\!n_{2}+g/n_{2}$. If $\widetilde{n}_{1}$ takes on the special value $\widetilde{n}_{1}\!=\!n_{1}+n_{2}$, then $g\!=\!n_{1}n_{2}$, which leads to $\widetilde{n}_{2}\!=\!n_{2}n_{2}\!=\!\widetilde{n}_{1}$. In other words, in this special case, changing the cone angle does not change $\Delta y$. We denote this special case the threshold $\widetilde{n}_{\mathrm{th}}\!=\!n_{1}+n_{2}$, which separates two regimes of the change in $\Delta y$ with $n$.

\end{enumerate}

\begin{figure*}[t!]
\centering
\includegraphics[width=9.6cm]{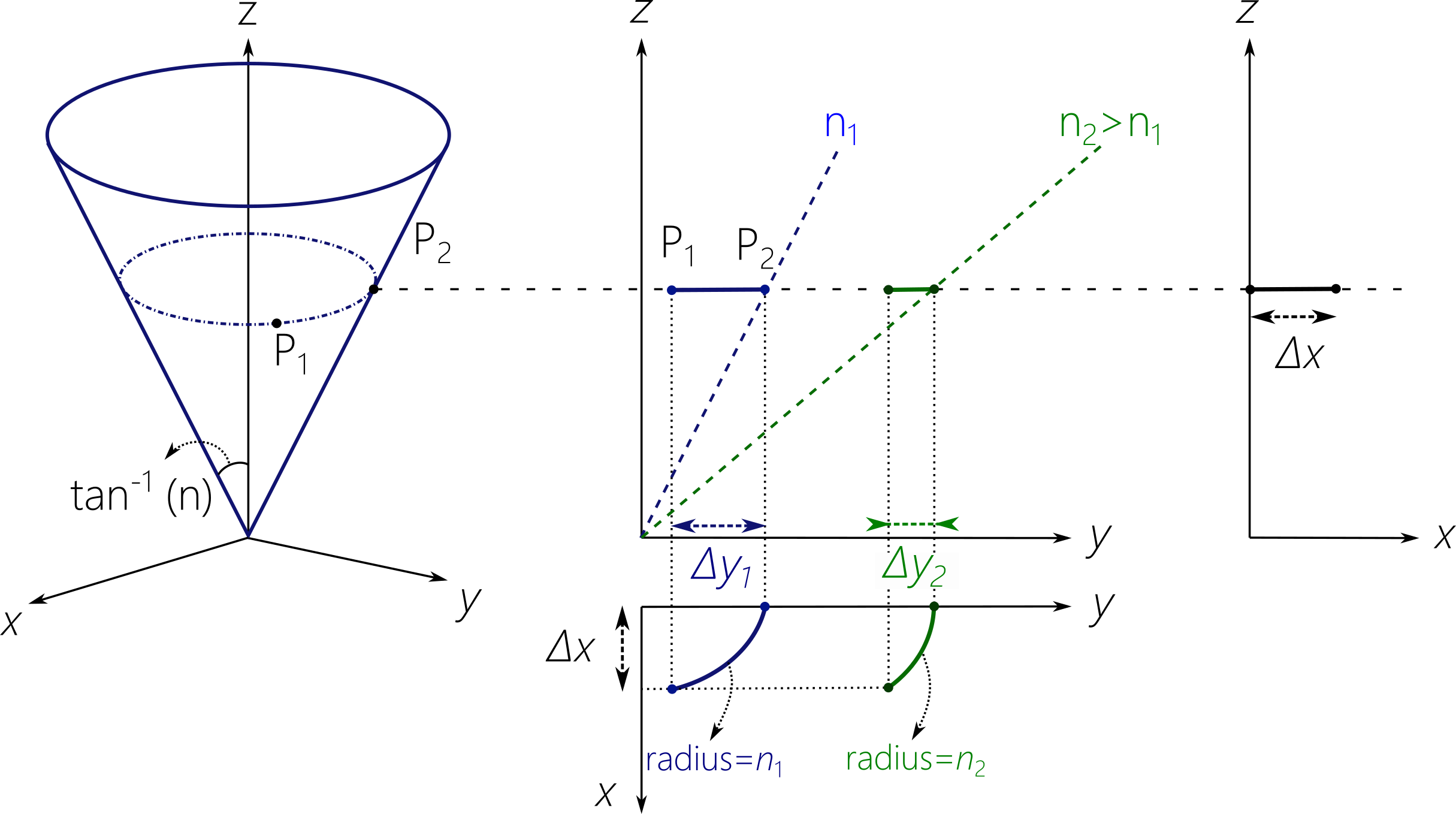}
\caption{Impact of the increase in the cone angle on two points P$_1$ and P$_2$ located at the same height when their $x$-coordinates and $z$-coordinates are fixed.}
\label{Fig:GenericConeSameHeight}
\end{figure*}

\begin{figure*}[t!]
\centering
\includegraphics[width=9.6cm]{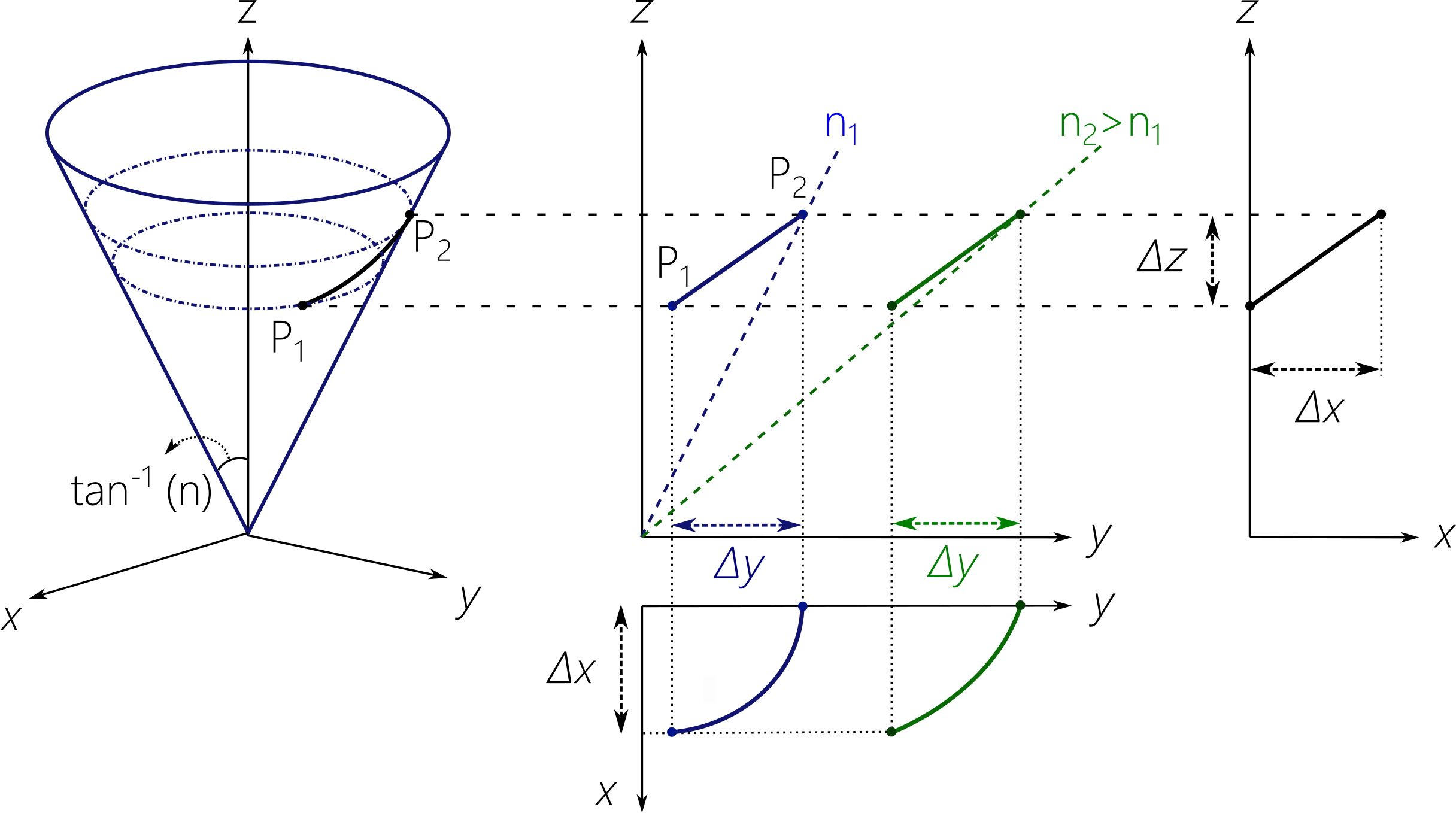}
\caption{Impact of the increase in the cone angle on two arbitrary points P$_1$ and P$_2$ when their $x$-coordinates and $z$-coordinates are fixed.}
\label{Fig:GenericConeGeneralPoints}
\end{figure*}

Consider two cones with $n_{2}\!>\!n_{1}$. When $\widetilde{n}_{1}\!=\!\widetilde{n}_{\mathrm{th}}+\delta\widetilde{n}$, then $\widetilde{n}_{2}\!=\!\widetilde{n}_{\mathrm{th}}+\tfrac{n_{1}}{n_{2}}\delta\widetilde{n}$. Therefore, when $\delta\widetilde{n}$ is positive we have $\widetilde{n}_{1}\!>\!\widetilde{n}_{2}$; i.e., the increase in $n$ led to a decrease in $\widetilde{n}$. On the other hand, when $\delta\widetilde{n}$ is negative we have $\widetilde{n}_{1}\!<\!\widetilde{n}_{2}$; i.e., the increase in $n$ led to an increase in $\widetilde{n}$. These two regimes correspond to the `anomalous' and `normal' refraction regimes in the main text. 

\section{Geometric interpretation of the law of refraction for ST wave packets at normal incidence}

\begin{figure*}[b!]
\centering
\includegraphics[width=11.6cm]{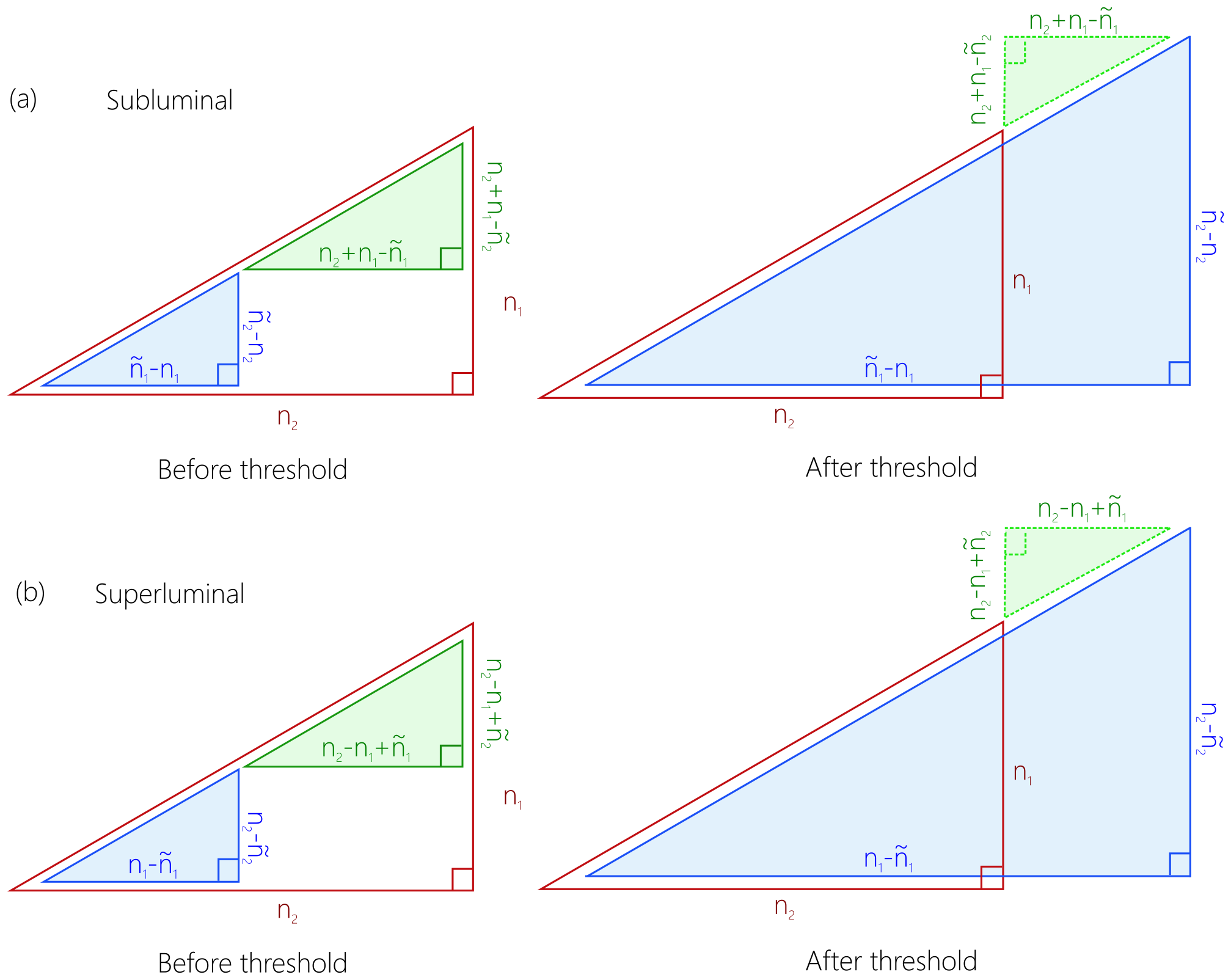}
\caption{Geometric representation of the law of refraction in terms of triangle similarities.}
\label{Fig:GeometricInterpretationTriangles}
\end{figure*}

The law of refraction for ST wave packets at normal incidence $n_{1}(n_{1}-\widetilde{n}_{1})\!=\!n_{2}(n_{2}-\widetilde{n}_{2})$ can be expressed in ratio form as follows,
\begin{equation}
\frac{n_{1}}{n_{2}}=\frac{n_{2}-\widetilde{n}_{2}}{n_{1}-\widetilde{n}_{1}}<1,
\end{equation}
where we assume that the ST wave packet traverses an interface between low-index and high-index media, $n_{1}\!<\!n_{2}$. Interpreting the ratio as the tangent of an angle $\psi\!<\!\tfrac{\pi}{4}$, $\tan{\psi}\!=\!\tfrac{n_{1}}{n_{2}}$, we construct a right-angled triangle with side lengths $n_{1}$ and $n_{2}$, as shown in Fig.~\ref{Fig:GeometricInterpretationTriangles}. The second ratio thus corresponds to a second triangle \textit{similar} to the first, in the sense of geometric similarity: the corresponding angles are equal and the corresponding sides have the same ratio to each other. In the subluminal regime $\widetilde{n}_{1}\!>\!n_{1}$ and $\widetilde{n}_{2}\!>\!n_{2}$, and the geometric construction is shown in Fig.~\ref{Fig:GeometricInterpretationTriangles}a. The threshold $\widetilde{n}_{1}\!=\!\widetilde{n}_{2}\!=\!n_{1}+n_{2}$ arises naturally when the two triangles coincide. In the superluminal regime $\widetilde{n}_{1}\!<\!n_{1}$ and $\widetilde{n}_{2}\!<\!n_{2}$, the geometric construction is shown in Fig.~\ref{Fig:GeometricInterpretationTriangles}b.

\section{Measurement Methodology}

\subsection{ST wave packet refraction through a layer}

Starting with a generic pulsed laser, the beam is split into two paths. In one path the ST wave packet is synthesized via a 2D pulse shaper that inculcates programmable spatio-temporal spectral correlations via a spatial light modulator to realize any spectral tilt angle $\theta_{1}$. The initial pulse is also utilized as a reference and traverses a second arm containing a delay. The ST wave packet and the reference pulse are then superposed and detected by an axially translatable CCD camera. When the two wave packets overlap in space and time, high-visibility spatially resolved fringes are observed. Placing a layer of thickness $L$ of a material having index $n_{2}$ in the common path in an ambient environment of index $n_{1}$ results in a loss of interference, but the fringes are recovered by adding an appropriate delay length $\Delta\ell$ in the reference path such that
\begin{equation}
(\widetilde{n}_{2}-n_{2})\!=\!(\widetilde{n}_{1}-n_{1})+n_{1}\Delta\ell/L.
\end{equation}
This formula can be rewritten in terms of measurable quantities,
\begin{equation}\label{Eq:DelayInSingleLayer}
\tau_{\mathrm{m}}=(n_{2}-n_{1})\tau_{\mathrm{o}}+\tau_{\mathrm{a}}+\tau_{\mathrm{d}},
\end{equation}
from which we can obtain $\theta_{2}$ and $\widetilde{v}_{2}$; here $\tau_{\mathrm{m}}\!=\!\widetilde{n}_{2}L/c$ and $\tau_{\mathrm{a}}\!=\!\widetilde{n}_{1}L/c$ are the group delays of the ST wave packet traversing a distance $L$ in the material and in free space, respectively, $\tau_{\mathrm{o}}\!=\!L/c$ is the delay of a traditional pulse in free space, and $\tau_{\mathrm{d}}\!=\!n_{1}\Delta\ell/c$ is the reference pulse delay. This approach can be generalized to the case of two material layers.

\subsection{ST wave packet refraction through bilayers}

Consider two layers of thicknesses $\ell_{1}$ and $\ell_{2}$ and indices $n_{1}$ and $n_{2}$, respectively, where $L\!=\!\ell_{1}+\ell_{2}$, surrounded symmetrically with a semi-infinite material of index $n_{0}$. First, assume only the first layer is placed in the common path of the ST wave packet and the reference pulse. The group delay balance implies that
\begin{equation}
\tau_{\mathrm{m}}^{(1)}=\tau_{\mathrm{a}}^{(1)}+\tau_{\mathrm{d}}^{(1)}+(n_{1}-1)\tau_{\mathrm{o}}^{(1)},
\end{equation}
where $\tau_{\mathrm{m}}^{(1)}\!=\!\ell_{1}\widetilde{n}_{1}/c$ is the group delay in the layer, $\tau_{\mathrm{a}}^{(1)}\!=\!\ell_{1}\widetilde{n}_{0}/c$ is the group delay in free space over a distance equal to the thickness of the layer, $\tau_{\mathrm{d}}^{(1)}\!=\!\Delta\ell^{(1)}/c$ is the delay in the reference arm, and $\tau_{\mathrm{o}}^{(1)}\!=\!\ell_{1}/c$ is the group delay of the reference pulse in free space over a distance equal to the thickness of the layer. Here $\widetilde{n}_{0}$ and $\widetilde{n}_{1}$ are the group indices in the ambient material and in the layer, which are related through the law of refraction $n_{0}(n_{0}-\widetilde{n}_{0})\!=\!n_{1}(n_{1}-\widetilde{n}_{1})$. A similar relationship holds when only the second layer is placed in the common path,
\begin{equation}
\tau_{\mathrm{m}}^{(2)}=\tau_{\mathrm{a}}^{(2)}+\tau_{\mathrm{d}}^{(2)}+(n_{2}-1)\tau_{\mathrm{o}}^{(2)},
\end{equation}
where the corresponding quantities for the second layer are $\tau_{\mathrm{m}}^{(2)}\!=\!\ell_{2}\widetilde{n}_{2}/c$, $\tau_{\mathrm{a}}^{(2)}\!=\!\ell_{2}\widetilde{n}_{0}/c$, $\tau_{\mathrm{d}}^{(2)}\!=\!\Delta\ell^{(2)}/c$, and $\tau_{\mathrm{o}}^{(2)}\!=\!\ell_{2}/c$. Here $\widetilde{n}_{0}$ and $\widetilde{n}_{2}$ are the group indices in the ambient material and in the layer, which are related through the law of refraction $n_{0}(n_{0}-\widetilde{n}_{0})\!=\!n_{2}(n_{2}-\widetilde{n}_{2})$.

When both layers are placed together in the common path and interference effects can be ignored, we can add the delays,
\begin{eqnarray}
\tau_{\mathrm{m}}\!\!\!\!&=&\!\!\!\!\tau_{\mathrm{m}}^{(1)}+\tau_{\mathrm{m}}^{(2)}\nonumber\\
\!\!\!\!&=&\!\!\!\!\tau_{\mathrm{a}}^{(1)}+\tau_{\mathrm{a}}^{(2)}+\tau_{\mathrm{d}}^{(1)}+\tau_{\mathrm{d}}^{(2)}+(n_{1}-1)\tau_{\mathrm{o}}^{(1)}+(n_{2}-1)\tau_{\mathrm{o}}^{(2)}\nonumber\\
\!\!\!\!&=&\!\!\!\!\tau_{\mathrm{a}}+\tau_{\mathrm{d}}-\tau_{\mathrm{o}}+\left(\frac{\ell_{1}}{L}n_{1}+\frac{\ell_{2}}{L}n_{2}\right)\tau_{\mathrm{o}},
\end{eqnarray}
where the quantities in this equation are defined as follows: 
\begin{eqnarray}
\tau_{\mathrm{a}}\!\!\!\!&=&\!\!\!\!\tau_{\mathrm{a}}^{(1)}+\tau_{\mathrm{a}}^{(2)}=\widetilde{n}_{0}\frac{L}{c},\\
\tau_{\mathrm{o}}\!\!\!\!&=&\!\!\!\!\tau_{\mathrm{o}}^{(1)}+\tau_{\mathrm{o}}^{(2)}=\frac{L}{c},\\
\tau_{\mathrm{d}}\!\!\!\!&=&\!\!\!\!\tau_{\mathrm{d}}^{(1)}+\tau_{\mathrm{o}}^{(2)}=\frac{\Delta\ell^{(1)}}{c}+\frac{\Delta\ell^{(2)}}{c}=\frac{\Delta\ell}{c}.
\end{eqnarray}

Note that of course the order of the two layers does not matter. The law of refraction for ST wave packets indicates that $n_{0}(n_{0}-\widetilde{n}_{0})\!=\!n_{1}(n_{1}-\widetilde{n}_{1})\!=\!n_{2}(n_{2}-\widetilde{n}_{2})$, such that the group index can be found in terms if the group index in the external medium,
\begin{eqnarray}
\widetilde{n}_{1}\!\!\!\!&=&\!\!\!\!n_{1}+\frac{n_{0}}{n_{1}}(\widetilde{n}_{0}-n_{0}),\\
\widetilde{n}_{2}\!\!\!\!&=&\!\!\!\!n_{2}+\frac{n_{0}}{n_{2}}(\widetilde{n}_{0}-n_{0}).
\end{eqnarray}

\subsection{Bilayers with zero group index}

Consider the bilayer scenario where the indices of the two layers are $n_{1}$ and $n_{2}$ and the thicknesses are equal. We aim to arrange for the group index in the first layer to be $\widetilde{n}_{1}\!=\!n_{1}-n_{2}$, which implies that the group index in the second layer is $\widetilde{n}_{2}\!=\!n_{2}-n_{1}\!=\!-\widetilde{n}_{1}$. By implementing the law of refraction at the interface between the external medium and the first layer, we obtain
\begin{equation}
\widetilde{n}_{0}=n_{0}-\frac{n_{1}n_{2}}{n_{0}}.
\end{equation}
When the external medium is free space $n_{0}\!=\!1$, we have $\widetilde{n}_{0}\!=\!1-n_{1}n_{2}$. Taking the indices of MgF$_2$, BK7, and sapphire to be 1.3751, 1.5108, and 1.7606, respectively, the requires spectral tilt angle $\theta_{0}$ required for various combinations are: (MgF$_2$,BK7)~$\rightarrow\theta_{0}\!=\!137.1^{\circ}$, (MgF$_2$,sapphire)~$\rightarrow\theta_{0}\!=\!144.86^{\circ}$, and (BK7,sapphire)~$\rightarrow\theta_{0}\!=\!148.9^{\circ}$. In all these configurations, the total group delay for the ST wave packet to traverse the bilayer is zero.

If $\widetilde{n}_{1}\!=\!n_{1}+n_{2}$, which implies that the group index in the second layer is $\widetilde{n}_{2}\!=\!n_{2}+n_{1}\!=\!\widetilde{n}_{1}$. By implementing the law of refraction at the interface between the external medium and the first layer, we obtain
\begin{equation}
\widetilde{n}_{0}=n_{0}+\frac{n_{1}n_{2}}{n_{0}}.
\end{equation}
When the external medium is free space $n_{0}\!=\!1$, we have $\widetilde{n}_{0}\!=1+\!n_{1}n_{2}$. Taking the indices of MgF$_2$, BK7, and sapphire to be 1.3751, 1.5108, and 1.7606, respectively, the requires spectral tilt angle $\theta_{0}$ required for various combinations are: (MgF$_2$,BK7)~$\rightarrow\theta_{0}\!=\!18^{\circ}$, (MgF$_2$,sapphire)~$\rightarrow\theta_{0}\!=\!16.29^{\circ}$, and (BK7,sapphire)~$\rightarrow\theta_{0}\!=\!15.28^{\circ}$. In all these configurations, the group delay for the ST wave packet is equal in both amplitude and sign in the two layers.

\section{Experimental details}

\begin{figure*}[b!]
\centering
\includegraphics[width=11.6cm]{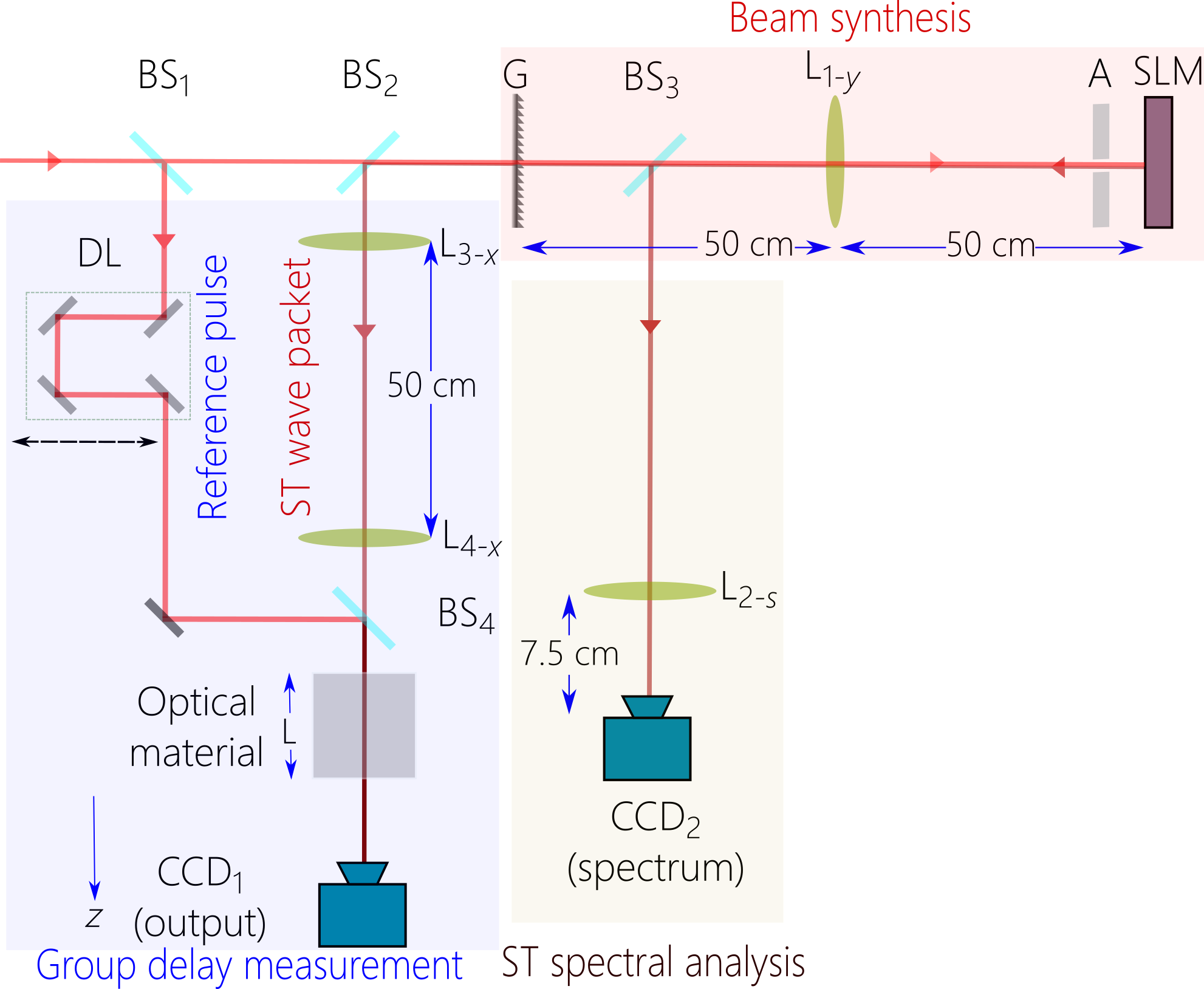}
\caption{Schematic of the optical arrangement for synthesizing and characterizing ST wave packets, and measuring group delays upon traversing material samples.}
\label{Fig:DetailedSetup}
\end{figure*}

We present in Fig.~\ref{Fig:DetailedSetup} a schematic of the setup used in our experiments. More details on the synthesis of ST wave packets using this approach can be found in Refs.~\cite{Kondakci17NP,Kondakci18PRL,Kondakci18OE,Kondakci18OL,Bhaduri18OE,Kondakci19ACSP}, and details of the interferometric approach to estimating the group delay in Refs.~\cite{Kondakci19NC,Bhaduri19Optica,Yessenov19OE}

A horizontally polarized pulsed laser beam from a Ti:Sapphire femtosecond laser (Tsunami, Spectra Physics) having a bandwidth of $\sim\!8.5$~nm and a central wavelength of $\sim\!800$~nm (pulse width $\sim\!100$~fs) was the source of pulses utilized in our experiments. The beam is first expanded spatially and collimated to produce a 25-mm-diameter plane-wave front. A reflective diffraction grating G (1200~lines/mm, area $25\times25$~mm$^2$; Newport 10HG1200-800-1) disperses the spectrum in space, and the second diffraction order is directed to a reflective spatial light modulator (SLM; Hamamatsu X10468-02) through a cylindrical lens L$_{1-y}$ having a focal length of $f\!=\!50$~cm. The SLM imparts a 2D phase pattern $\Phi(x,y)$ to jointly modulate the spatial and temporal spectra. An aperture A is introduced in front of the SLM to reduce the temporal bandwidth to 0.2~nm. The modulated wave front is then retro-reflected back to the grating through the same lens L$_{1-y}$ to reconstitute the pulse. The spatial frequencies are simultaneously overlapped to produce the ST light sheet in the same step.

To measure the spatio-temporal spectrum, we sample a portion of the retro-reflected field from the SLM after passing through the lens L$_{1-y}$ via a beam splitter BS$_{3}$ as shown in Fig.~\ref{Fig:DetailedSetup}. The field is directed through a spherical lens L$_{s-4}$ having $f\!=\!7.5$~cm to a CCD camera (CCD$_2$; the Imaging Source, DMK 72AUC02) in a $2f$ configuration to collimate the spectrum along the $y$-direction and carry out a Fourier transform spatially along the $x$-direction.

The reference pulse is obtained from the initial laser pulses via the beam splitter BS$_1$. A neutral density filter adjusts the power power, and the beam is spatially filtered, expanded, and collimated using two spherical lenses ($f\!=\!50$~cm and 10~cm) and a pinhole (diameter 30~$\mu$m) in a $4f$ configuration with the pinhole at the Fourier plane. An optical delay line is introduced in the reference path to adjust the path difference with the ST wave packet arm before combining the two beams via the beam splitter BS$_4$. The resulting interferogram is recorded with a CCD camera (CCD$_1$; The Imaging Source, DMK 33UX178). The glass samples (in the form of flat windows) are placed in the common path after BS$_4$ and the CCD$_1$ is mounted on a translation stage (Thorlabs LTS150) to move along the common path.

\subsection{Measuring the group velocity of the ST wave packet}

The maximum fringe visibility in the interference of wave packets occurs when the optical path difference between the ST wave packet (pulse width $\sim\!9$~ps) and reference pulse (pulse width $\sim\!100$~ps) is close to zero at the time they reach the detector, thereby indicating that the two wave packets overlap in space and time. When the detector in the common path (free space) is moved a distance $L$, the optical path lengths for both ST wave packet and reference pulse change, which results in a loss of fringe visibility. The spatially resolved high-visibility fringes are regained by introducing a delay distance $\Delta\ell$ in the reference path using a delay line (Thorlabs DDS300) when a new temporal balance is reached between the delays.

Similarly, when a material of length $L$ and index $n$ is introduced into the common path, the optical path lengths of ST wave packet and the reference pulse are changed, which diminishes the fringe visibility. However, adding an extra free-space delay length $\Delta\ell$ in the reference path can restore the temporal balance of the ST wave packet and reference pulse, and high-visibility fringes reappear. Finally the group velocity or group delay is estimated using the length $L$ and $\Delta\ell$ (Eq.~\ref{Eq:DelayInSingleLayer}).

\subsection{Measurement of temporal envelope of ST wave packets}

\begin{figure*}[b!]
\centering
\includegraphics[width=14.6cm]{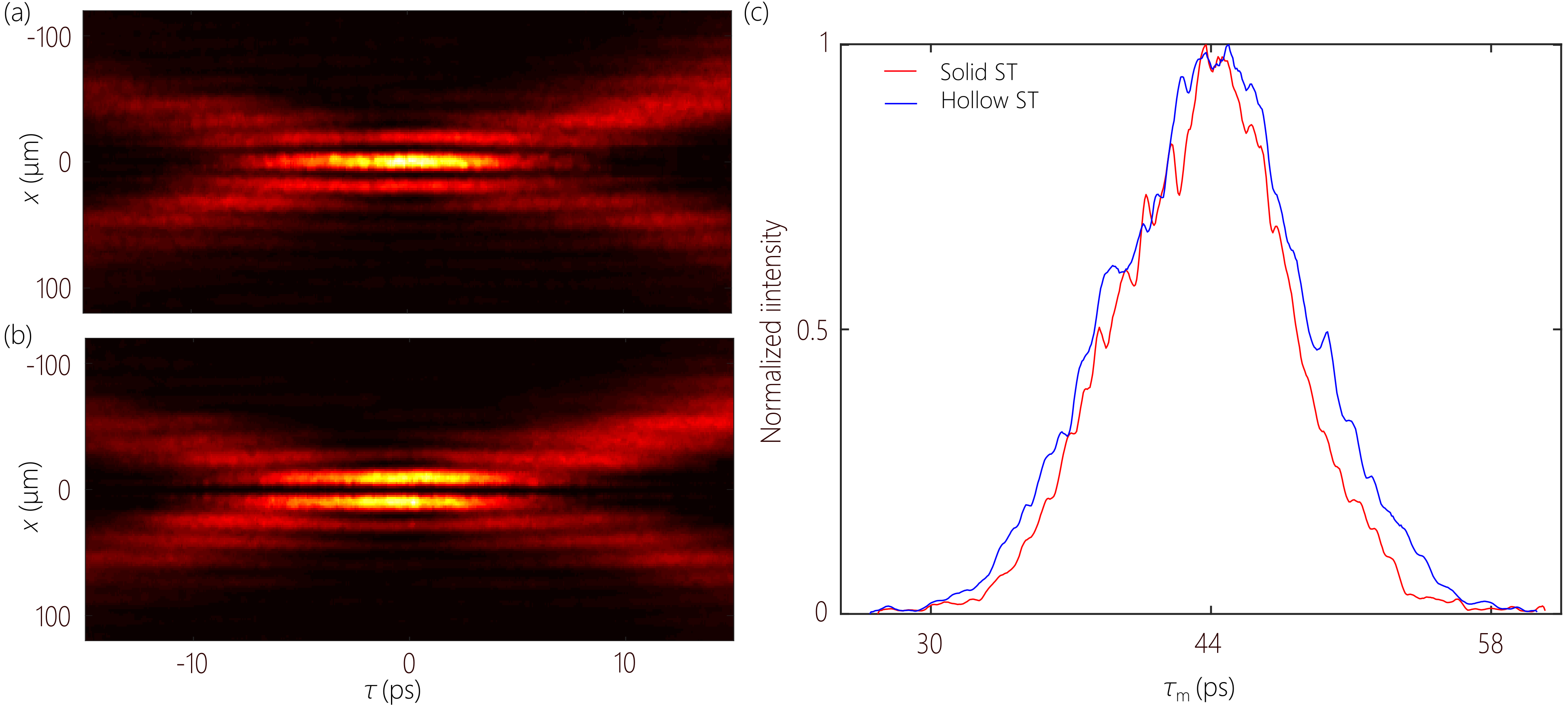}
\caption{Measured spatio-temporal intensity profile $I(x,\tau)$ of ST wave packets; central wavelength is $\lambda_{\mathrm{o}}\!\approx\!797$~nm, $\theta\!=\!70^{\circ}$, bandwidth is $\Delta\lambda\!\approx\!0.2$~nm, spectral uncertainty is $\delta\lambda\!\approx\!30$~pm, beam width is $\Delta x\!\approx\!14$~$\mu$m, and pulse width $\Delta\tau\!\approx\!9.5$~ps. (a) Measured profile $I(x,z\!=\!0,\tau)$ for a ST wave packet with the parameters listed above and all the frequency components in phase, leading to a peak at the center of the profile. (b) Same as (a), but with a $\pi$ phase introduced between the positive and negative spatial frequencies, leading to a null at the profile center. The change in the profile does not affect the refractory properties of the ST wave packet. (c) The temporal profile $I(0,0,\tau)$ at $x\!=\!0$ for the ST wave packets in (a) and (b).}
\label{Fig:STWavePacket}
\end{figure*}

For any given value of the spectral tilt angle $\theta$, the spatio-temporal intensity profile of the ST wave packet is obtained by fixing the location of CCD$_1$ and recording the interference pattern while varying the reference path delay $\Delta\ell$ after centering the delay value that maximized the fringe visibility. The detector captures 15 interferograms at each value of $\Delta\ell$. The delay line is advanced in increments of 10~$\mu$m, with a total of 201 steps (100 steps on either side of the zero optical path difference). The same approach is followed when a sample is placed in the common path.

We plot in Fig.~\ref{Fig:STWavePacket}a and Fig.~\ref{Fig:STWavePacket}b the measured spatio-temporal intensity profile of two ST wave packets. The wave packets share the same spectral tilt angle $\theta\!=\!70^{\circ}$, with the only difference being the change in the spatial profile. In Fig.~\ref{Fig:STWavePacket}a we have a peak in the center and in Fig.~\ref{Fig:STWavePacket}b we have instead a null produced by inserting a relative phase difference of $\pi$ between the positive and negative spatial frequency components. Despite the difference in spatial profile, the refractory properties of both these ST wave packets are the same and follow solely from $\theta$.

\begin{figure*}[t!]
\centering
\includegraphics[width=14.6cm]{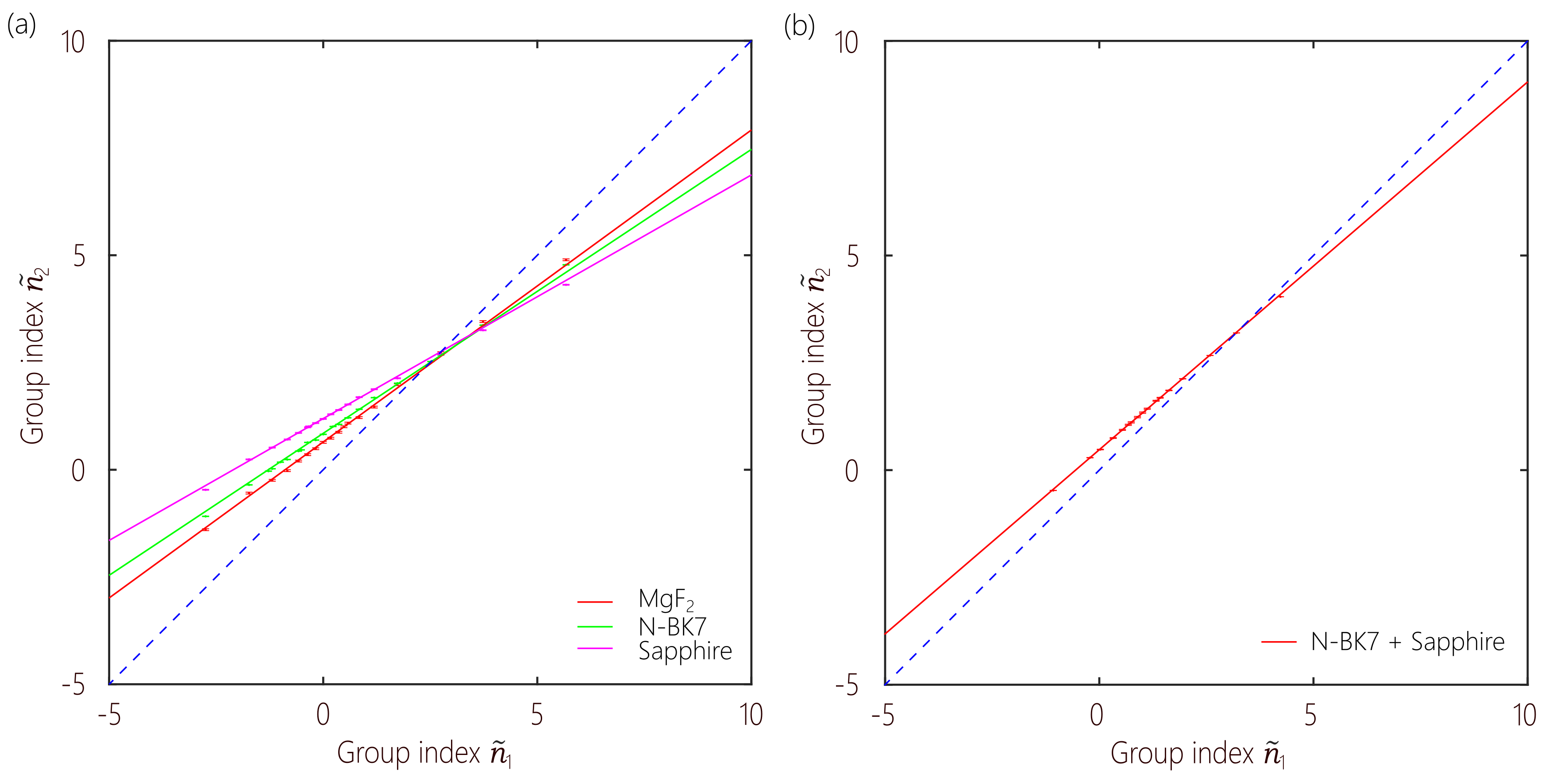}
\caption{Measurements verifying the law of refraction of ST wave packets at normal incidence plotted as a transformation between incident and transmitted group indices $\widetilde{n}_{1}$ and $\widetilde{n}_{2}$. (a) The data corresponds to that in Fig.~2a and (b) to that in Fig.~2b in the main text.}
\label{Fig:NormalIncidenceData}
\end{figure*}

We plot in Fig.~\ref{Fig:NormalIncidenceData} the data for the change in the group index when going from one medium to another, confirming the linear relationship between $\widetilde{n}_{1}$ and $\widetilde{n}_{2}$ predicted by the law of refraction of ST wave packets given in Eq.~1 of the main text.

\subsection{Measurements at oblique incidence}

For measurements at oblique incidence, the sample is mounted in such a way that it can be rotated around the center of its front surface, which intercepts with the width of the ST beam along the $x$-direction. Each point on the new curve representing the law of refraction of ST wave packets at oblique incidence is obtained from the change in the delay line $\Delta\ell$ required to retrieve high-visibility interference fringes after inserting the sapphire window at an inclined angle with respect to the maximum-visibility delay in absence of it. Using the measured values of $\Delta\ell$, the spectral tilt angles and group indices are calculated using the Eq.~\ref{Eq:DelayInSingleLayer} after replacing the sample thickness $L$ with $L/\cos{\phi_{2}}$. Note that the group velocity of the ST wave packet in the material changes with incident angle whereas that of the reference pulse does not. Using this approach, we confirm the variation in the refraction law of ST wave packets as the angle of incidence is changed.

The measurements of the change in the group index of the transmitted ST wave packet with incidence angle of ST beam [Fig.~4a in the main text] were carried out using a sapphire sample after holding fixed the incident group index (a subluminal value of $\theta_{1}\!=30^{\circ}$ and a superluminal value $\theta_{1}\!=\!108.7^{\circ}$) while varying the sample inclination $\phi_{1}$ with respect to the incident ST wave packet.

\clearpage
\bibliography{diffraction1}

\end{document}